\documentclass[%
 reprint,
 amsmath,amssymb,
 aps,
]{revtex4-2}

\usepackage{graphicx}
\usepackage{dcolumn}
\usepackage{bm}
\usepackage{placeins}
\usepackage{physics}
\usepackage{siunitx}
\usepackage{bullcntr}
\usepackage{xcolor}

\begin{document}

\preprint{APS/123-QED}

\title{Impacts of Random Filling on Spin Squeezing via Rydberg Dressing in Optical Clocks}

\author{Jacques Van Damme}%
\author{Xin Zheng}%
\author{Mark Saffman}%
\author{Maxim G. Vavilov}%
\author{Shimon Kolkowitz}
\email{kolkowitz@wisc.edu}

\affiliation{%
 Department of Physics, University of Wisconsin - Madison, Madison, WI 53706
}

\date{\today}


\begin{abstract}
We analyze spin squeezing via Rydberg dressing in optical lattice clocks with random fractional filling.  We compare the achievable clock stability in different lattice geometries, including unity-filled tweezer clock arrays and fractionally filled lattice clocks with varying dimensionality. We provide practical considerations and useful tools in the form of approximate analytical expressions and fitting functions to aid in the experimental implementation of Rydberg-dressed spin squeezing. 
We demonstrate that spin squeezing via Rydberg dressing in one-, two-, and three-dimensional optical lattices can provide significant improvements in stability in the presence of random fractional filling. 

\end{abstract}

\maketitle



\section{\label{sec:introduction}Introduction}
Since their advent roughly seven decades ago, atomic clocks have developed at a rapid pace. State-of-the-art optical lattice clocks are now operating at or near the standard quantum projection noise (QPN) limit ~\cite{atom_clock,ab4089,marti2018imaging,oelker2019demonstration}, particularly in differential clock comparisons where correlated noise spectroscopy can be employed to remove local oscillator noise~\cite{hume2016probing,marti2018imaging,tan2019suppressing,young2020tweezer,clements2020lifetime}. Extremely precise clock comparisons can be used for exciting applications such as tests of relativity \cite{relativity}, searches for dark matter \cite{counts2020evidence}, tracking of space-craft \cite{RevModPhys.87.637}, and relativistic geodesy~\cite{geodesy}. Further improvements in clock stability will enhance their use for these applications, and will open the door to new applications in searches for beyond Standard Model physics and gravitational wave detection~\cite{new_physics,gravitationalwaves}. 

The QPN limit arises from the independent projection onto the clock state basis of each of the atoms used to reference the local oscillator. Introducing entanglement between the atoms in the form of spin squeezing can reduce the uncertainty associated with these measurements~\cite{science.1104149,RevModPhys.90.035005,PhysRevA.47.5138}. A high degree of spin-state squeezing in large atom ensembles and optical lattice clocks has been demonstrated using coupling of the atoms to an optical cavity~\cite{cavitysqueezing,2006.07501}. Recently, an attractive alternative approach was proposed in which the squeezing is generated through dressed Rydberg interactions that can be turned on and off with one additional laser~\cite{mainpaper}, and there have been subsequent demonstrations of the one-axis twisting Hamiltonian using dressed Rydberg interactions in alkali atoms~\cite{PhysRevLett.124.063601}, as well as the generation of entangled Bell states in pairs of strontium atoms using the same transitions~\cite{madjarov2020high}. However, spin-squeezing of an optical clock transition via Rydberg-dressing has yet to be experimentally demonstrated, and there are a number of important practical considerations that have not previously been investigated.

There are now three different geometric configurations of neutral atom optical clocks. In most current optical lattice clocks a roughly $10^3-10^4$ site one-dimensional lattice is loaded stochastically from a magneto-optical trap (MOT). The resulting random variations in the number of atoms occupying each site can result in dephasing due to on-site atom-atom interactions at high atom densities \cite{nicholson2015systematic} or a lattice filling fraction $P_{\textrm{filling}}<1$ in the low-density regime. A recently demonstrated alternative approach makes use of evaporative cooling to load Fermi-degenerate atoms into a three-dimensional lattice with a filling fraction approaching unity \cite{campbell2017fermi,marti2018imaging,oelker2019demonstration}. Finally, there have been recent demonstrations of tweezer array clocks, in which roughly 10-100  individual dipole tweezer traps are loaded into an array with either zero or one atom on each site~\cite{PhysRevX.9.041052,Endres1024,young2020tweezer}. The array can then be reconfigured to deterministically realize one atom per site ($P_{\textrm{filling}}=1$) \cite{Endres1024}, or the use of variational quantum algorithms has been proposed to adjust the dynamics on the fly to optimize the generation of spin-squeezing for a given random configuration \cite{kaubruegger2019}. 

In this work we investigate the impact of fractional filling on the degree of spin-squeezing that can be achieved using Rydberg-dressing in the various geometric configurations of optical lattice clocks. We find that fractional filling reduces the achievable degree of squeezing, primarily by increasing the average inter-atomic spacing, with disorder playing only a small role. We consider the impact of random fractionally-filled configurations on the optimal squeezing time and angle of minimal uncertainty, and find that useful levels of squeezing can still be achieved in fractionally filled one-dimensional optical lattices. Finally, we compare the limits on achievable spin-squeezed clock stabilities for the different clock geometries as a function of the number of atoms. 

\begin{figure}[ht]
    \centering
    \includegraphics[page=1,width=0.45\textwidth]{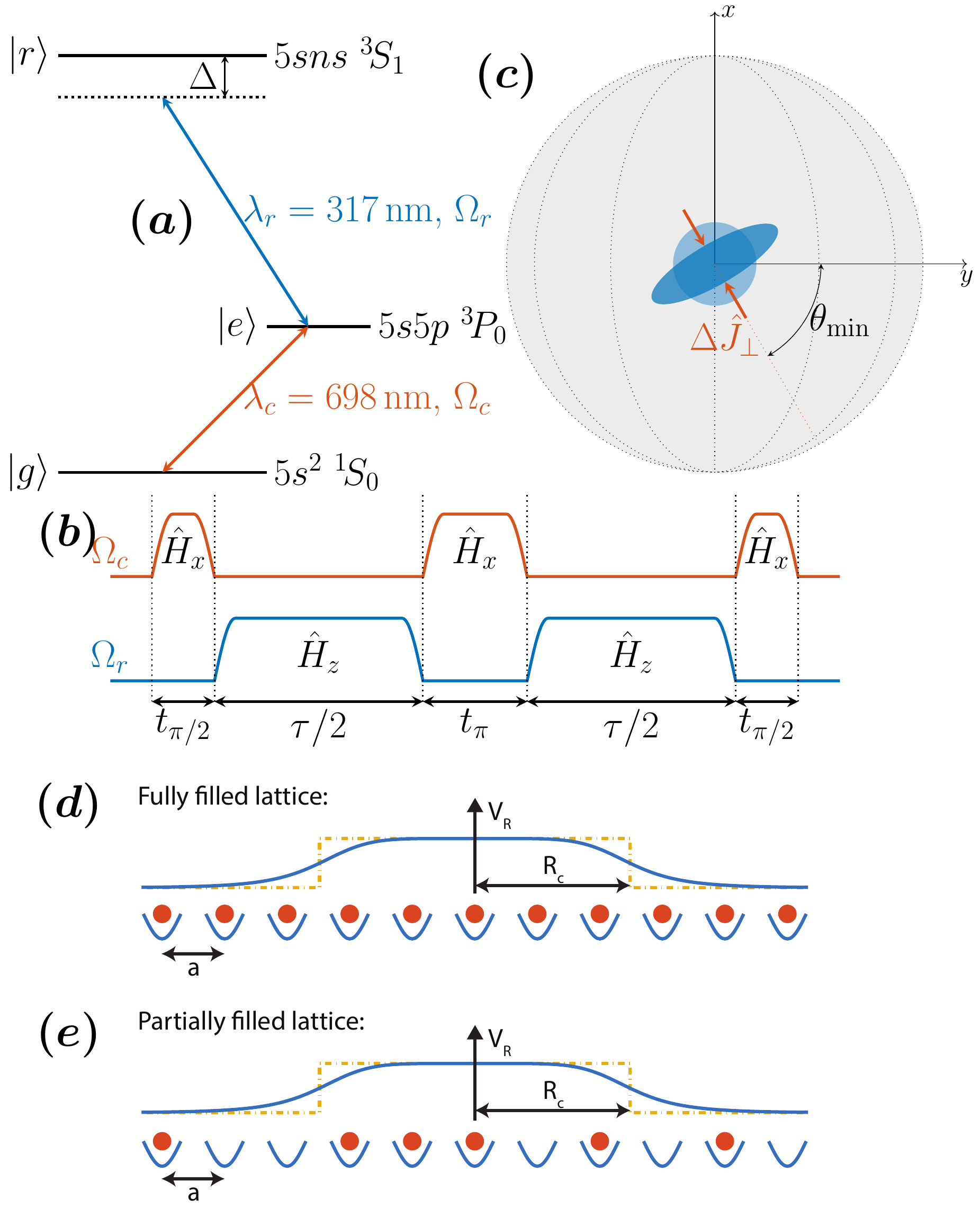}
    \caption{(a) Strontium energy level diagram. The clock laser drives the transition between $\ket{g}$ and $\ket{e}$ with Rabi frequency $\Omega_c$. The squeezing laser drives the Rydberg transition between $\ket{e}$ and $\ket{r}$ far off-resonance with Rabi frequency $\Omega_r$.
    (b) Spin-echo sequence of rotation (with the clock transition) and squeezing intervals (with the Rydberg transition) to perform one-axis squeezing of the coherent spin state $\ket{g}^{\otimes N}=\ket{-1/2}^{\otimes N}$. (c) Bottom view of the Bloch sphere illustrating the squeezing of the binomial distribution uncertainty of the coherent spin state towards reduced uncertainty $\Delta\hat{J}_{\perp}$ along the perpendicular direction at angle $\theta_{\textrm{min}}$.(d),(e) Fully or partially filled one-dimensional lattice with the dressed Rydberg interaction potential given by Eq.~\eqref{eq:Vij} for the central atom (solid line) and the corresponding Heaviside approximation (dashed line).
    }
    \label{fig:intro}
\end{figure}

\section{Spin Squeezing via Rydberg Dressing}

We analyze the  method for generating spin squeezing on the strontium optical clock transition using Rydberg dressing~\cite{mainpaper} as illustrated in Fig.~\ref{fig:intro}. Figure~\ref{fig:intro}(a) shows the relevant energy level diagram for strontium. In order to simplify our analysis in this work we consider only the stable isotopes of strontium with zero nuclear spin (${}^{88}$Sr, ${}^{86}$Sr,${}^{84}$Sr). In an applied magnetic field, the doubly-forbidden clock transition between $\ket{g}$ and $\ket{e}$ can be driven with the clock laser and used for Rabi rotations of the effective two-level atoms. In this effective spin-1/2 picture we define $\ket{g}=\ket{-1/2}$ and $\ket{e}=\ket{+1/2}$ with spin operators $\hat{J_{\alpha}} = \sum_{i}\hat{\sigma}_{\alpha}^{(i)}$, where $\hat{\sigma}_{\alpha}^{(i)}$ is the Pauli operators $\alpha \in \{x,y,z\}$ acting on atom $i$. The Rydberg transition between $\ket{e}$ and $\ket{r}$ is driven far off-resonance with one additional Rydberg laser $\Omega_r$. In this regime the frequency detuning $\Delta$ of the Rydberg laser from the resonant transition is large,  $|\Delta| \gg \Omega_r$, and the Rydberg states are only virtually excited and perturbation theory up to fourth order introduces an energy shift on the $\ket{e}$ energy level that includes an approximate interaction potential between pairs of atoms in the lattice $V_{ij}$~\cite{mainpaper,PhysRevA.82.033412,PhysRevA.94.023408,zeiher2016,PhysRevLett.104.195302,PhysRevLett.124.063601,PhysRevA.65.041803}:

\begin{equation}
    V_{ij} = V(\boldsymbol{r_i},\boldsymbol{r_j}) = \frac{\hbar\Omega_r^4}{8|\Delta|^3}\frac{R_c^6}{|\boldsymbol{r_i}-\boldsymbol{r_j}|^6 + R_c^6}
    \label{eq:Vij}.
\end{equation}

\noindent This approximate potential has a finite range Rydberg interaction radius $R_c=\left|C_6/(2\hbar|\Delta|)\right|^{1/6}$ that spans several lattice sites, illustrated by the solid line 
above the one-dimensional lattice of Fig.~\ref{fig:intro}(d). The system is described by two Hamiltonians that can be controlled separately, $\hat{H}_x$ from the clock laser and $\hat{H}_z$ from the Rydberg laser:

\begin{subequations}
    \label{eq:H}
\begin{eqnarray}
        \hat{H}_x &= & \frac{\hbar\Omega_c}{2}\sum\limits_{i=1}^{N}\hat{\sigma}_x^{(i)} \, , \\
        \hat{H}_z &= & \sum_{i<j}^{N}\frac{V_{ij}}{4}\hat{\sigma}_z^{(i)}\hat{\sigma}_z^{(j)} + \sum_{i=1}^{N}\frac{\delta_i}{2}\hat{\sigma}_z^{(i)}\, ,
\end{eqnarray}
\end{subequations}

with $\delta_i$ the inhomogeneous longitudinal field contributions. 
 Using a spin-echo sequence with the clock laser and Rydberg laser according to
Fig.~\ref{fig:intro}(b) on the initial coherent spin state $\ket{g}^{\otimes N}=\ket{-1/2}^{\otimes N}$ removes this unwanted linear procession due to the $\hat{\sigma}_z$ terms, and results in squeezing via one-axis twisting from the effective atom-atom interaction~\cite{PhysRevA.47.5138,mainpaper}. 
After the spin-echo sequence, the atom ensemble forms a spin squeezed state with main spin direction along the z-axis and a squeezed uncertainty ellipse with minimal variance along the direction at angle $\theta_{\textrm{min}}$ illustrated in Fig.~\ref{fig:intro}(c). These spin squeezed states can then be used for clock interrogation along the direction of minimal uncertainty to beat the QPN limit. 

The squeezing parameter $\xi^2$ equals the ratio of the squeezed state variance to the coherent state variance, $\xi^2_{\textrm{min}}$ is the minimal squeezing parameter along the $\theta_{\textrm{min}}$ direction obtained with the optimal squeezing time $\tau_{\textrm{opt}}$.

\begin{equation}
\xi_{\rm min}^2 = \frac{N\langle\hat{J}^2_{\perp}(\theta_{\rm min})\rangle}{\langle \boldsymbol{\hat{J}} \rangle^2},\quad 
\hat{J}_{\perp}(\theta) = 
\cos(\theta)\hat{J}_x + \sin(\theta)\hat{J}_y
\label{eq:xi}
\end{equation}

\noindent In~\cite{mainpaper}, analytical solutions were derived for the parameters required to evaluate Eq.~\eqref{eq:xi} for the spin-squeezing sequence shown in Fig.~\ref{fig:intro}(b). We present them here:
\begin{subequations}
\label{eq:solution}
\begin{eqnarray}
	&&\langle\hat{J}_z\rangle  = -\frac{1}{2}\sum\limits_{i=1}^{N}\prod\limits_{k\neq i}^{N}\cos\left(\frac{V_{ij}\tau}{2\hbar}\right) \\
	&&\langle\hat{J}_x^2\rangle   =  \frac{N}{4} + \frac{1}{4}\sum\limits_{i<j}^{N}\Bigg[ \prod\limits_{k\neq i,j}^{N}
	\cos\left(\frac{(V_{ik}-V_{jk})\tau}{2\hbar}\right) \nonumber 
	\\
	&&- \prod\limits_{k\neq i,j}^{N}\cos\left(\frac{(V_{ik}+V_{jk})\tau}{2\hbar}\right)\Bigg]
	\\
	&&\langle\hat{J}_y^2\rangle = \frac{N}{4} \\
	&&\langle\hat{J}_x\hat{J}_y + \hat{J}_y\hat{J}_x\rangle  =-\sum\limits_{i<j}^{N}\sin\left(\frac{V_{ij}\tau}{2\hbar}\right)\prod\limits_{k\neq i,j}^{N}\cos\left(\frac{V_{ik}\tau}{2\hbar}\right) \nonumber
	\\
\end{eqnarray}
\end{subequations}
While the computation of these terms scales polynomially in the number of atoms in the lattice, it can still be computationally intensive for large lattices $\ge 10^3$. We have therefore derived exact analytical expressions for these terms in the approximation where the Rydberg interaction potential is replaced with a Heaviside function as illustrated by the dashed curve above the lattice in Fig.~\ref{fig:intro}(d),  that can be trivially computed for all lattice sizes $N \ge 2 R_c/a$, see  appendix~\ref{app:heaviside}. We note that this approximation was previously used in Ref.~\cite{PhysRevA.94.023408} (without an analytical expression), and as in that work we find that it is a good approximation. 

The widely used interaction potential in Eq.~\eqref{eq:Vij} is in fact also an approximation based on the assumption that the dipole-dipole interaction between two Rydberg states is dominated by the long range van der Waals contribution $V_{\rm vdW}\propto 1/r^6$. However in~\cite{Saffman_2016} the Rydberg dressing interaction is derived from the  dipole-dipole potential $V_{\rm dd}$ including intermediate range $1/r^3$ scaling, 
\begin{subequations}
\begin{eqnarray}
\label{eq:Vdd}
&&V_{\rm dd} = \frac{\delta}{2}-\frac{\delta}{2}\sqrt{1+\left(\frac{x_c}{r}\right)^6} \, ,\\
&&V_{\rm vdW} = \frac{-\delta}{4}\left(\frac{x_c}{r}\right)^6 = \frac{C_6}{r^6}\, ,
\end{eqnarray}
\end{subequations}

\noindent where $\delta$ is the F\"{o}rster defect, and $x_c$ the length scale for the transition from $1/r^3$ to $1/r^6$ scaling of the potential~\cite{PhysRevA.77.032723}. The corresponding dressed energy shift on the excited clock state $\ket{e}$ will include the exact $V_{ij}(r)$ Rydberg interaction potential presented in Eq.~\eqref{eq:Vij_exact}.
The quality of the long range van der Waals approximation is strongly dependent on the F\"{o}rster resonance $\delta$ (illustrated by Fig.~\ref{fig:xi2_Vij_exact} and Fig.~\ref{fig:tau_Vij_exact} in the appendix).

The F\"orster energy defects of the $^3\hspace{-.1cm}S_1$ 
Sr series can be estimated using quantum defect data from Refs.~\cite{Vaillant_2012,Ding2018}. We have assumed that coupling between triplet and singlet states is negligible and neglected any singlet component of the triplet series.  

The defect assuming coupling to $^3\hspace{-.1cm}P_J$ states was calculated from
\begin{equation}
\label{eq:J}
\hbar\delta_J(n)=U(n,^3\hspace{-.1cm}P_J)+U(n-1,^3\hspace{-.1cm}P_J)-2U(n,^3\hspace{-.1cm}S_1), 
\end{equation}

\noindent where the $U$ are term energies $U(n, ^{(2S+1)}\hspace{-.1cm}L_J)=-{E_H/2[n-\nu_J(n)]^2}$ and  $\nu_J(n)$ are the $n$ and $J$ dependent quantum defects. For $E_H$ we use the finite mass corrected value of the Hartree energy.

As can be seen in Fig.~\ref{fig:V(R)Mark} of appendix~\ref{app:Vij_exact}, all $J$ channels give a defect that is comparable in magnitude so it is sensible to characterize the interaction with a spin weighted effective value. 
The spin weighted value was calculated as 

\begin{equation}
\delta(n)=\frac{\sum_{J=0,1,2}(2J+1)\delta_J(n)}{\sum_{J=0,1,2}2J+1}.
\label{eq:delta}
\end{equation}

\noindent The $\delta=\SI{-1.26}{\giga\hertz}$ we calculate for the $5s55s~{}^3\hspace{-.1cm}S_1$ Rydberg level results in a potential that is well approximated by $V_{ij}$ of Eq.~\eqref{eq:Vij} and we use it for the rest of this work.

It should be noted that this discussion is directly  applicable to the bosonic isotopes of Sr which have zero nuclear spin. Rydberg interactions between atoms with hyperfine structure may excite resonances when the hyperfine splitting  of the ionic core matches the energy separation of Rydberg levels~\cite{Robicheaux2018}. Calculations that include the hyperfine 
structure~\cite{Robicheaux2019} in $^{87}$Sr predict  strong resonances leading to orders of magnitude increases in the $C_6$   coefficent at $n=64$ and $105$. Performing dressing near these values of $n$ may offer a route to  enhanced  values of $R_c/a$.\\

Prior works have argued that far off-resonance Rydberg dressing is advisable due to the improved lifetimes $\tilde{\tau} \propto \Delta^2$~\cite{mainpaper,PhysRevA.94.023408}. However, the ratio of the life time to squeezing time $\tilde{\tau}/\tau_{\textrm{opt}}$ (as well as the degree of squeezing) improves at lower detuning (as shown in appendix~\ref{app:detuning}). This means that the far off-resonance regime is not necessarily justified in terms of spin squeezing performance. This conclusion also follows from~\cite{Saffman_2016}, where a figure of merit for optimal detuning in terms of the ratio between the coherent operation time and state lifetime is defined for one-dimensional lattices. Nevertheless, as with prior works we stay within the far off-resonance regime with $\Delta = -10\Omega_r$ for this work because the resulting spin-squeezing can be treated analytically, and we leave an analysis of the closer-to-resonance case for future work.

\section{One-Dimensional Lattice Squeezing}

\subsection{Partial Lattice Filling}
Prior to this work the achievable spin squeezing via Rydberg dressing has only been studied for fully filled lattices \cite{mainpaper,PhysRevA.94.023408}, however this condition is not realistic for most optical lattice clocks. In standard one-dimensional optical lattice clocks, atoms are first trapped and cooled with a MOT and then loaded into the lattice potential. After this procedure multiple atoms are present at every lattice site. To protect the clock state from collisional decoherence, a photo-association step can be employed to kick out atoms in pairs from the lattice until only empty or singly occupied sites remain as illustrated by Fig.~\ref{fig:intro}(e). The expected filling probability of each lattice site is $P_{\textrm{filling}}=50\%$, and this fractional filling influences the spin squeezing. Similarly, in recent demonstrations of Fermi-degenerate 3D optical lattice clocks the fraction of sites with only one atom is less than unity due to the finite temperature of the atoms when loaded into the lattice \cite{campbell2017fermi,marti2018imaging}.   

\begin{figure*}[ht]
    \centering
    \includegraphics[page=2,width=\textwidth]{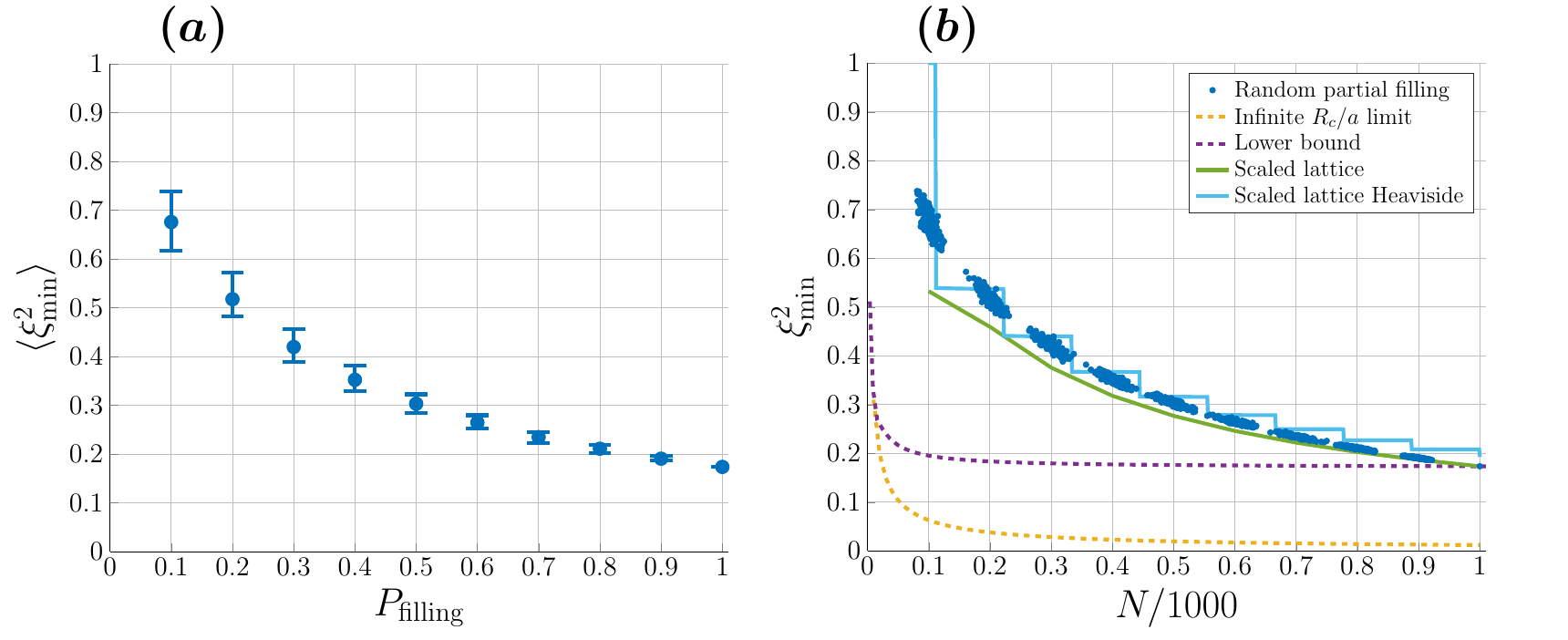}
    \caption{(a) The minimal squeezing parameter $\xi^2_{\textrm{min}}$ averaged over 200 randomly filled one-dimensional lattice configurations ($M=10^3$) at each lattice site filling probability $P_{\rm filling} \in \{0.1, 0.2, ..., 1\}$ with Rydberg interaction radius $R_c/a=9$. The error bars indicate the minimal and maximal value from the simulations.
    (b) Corresponding scatter plot of minimal squeezing $\xi^2_{\textrm{min}}$ plotted versus the actual filling fraction $N/1000$ loaded into the lattice.
    }
    \label{fig:xi2_vs_Pfilling}
\end{figure*}

In order to quantify the impact of random fractional filling, we performed simulations on random partially filled one-dimensional lattices with various lattice site filling probabilities $P_{\textrm{filling}}\in \{0.1,0.2,...,1\}$ using Eq.~\eqref{eq:solution} on a cluster computer. For every simulation, a lattice is randomly filled with the occupation probability of each site equal to $P_{\rm filling}$ and the interactions $V_{ij}$ involving empty sites are made zero in solution Eq.~\eqref{eq:solution}. The squeezing parameter is minimized in each individual lattice configuration by selecting the optimal squeezing time $\tau_{\rm opt}$ and angle of minimal uncertainty $\theta_{\rm min}$. During clock operation, these shot-to-shot optimizations for each randomly loaded lattice will likely not be practical, and we address this in section~\ref{sec:par_opt}. In Fig.~\ref{fig:xi2_vs_Pfilling}(a) the average squeezing parameter $\langle \xi^2_{\rm min} \rangle$ is plotted for each of the considered filling probabilities using the experimental parameters listed in Table~\ref{tab:table1}, with corresponding Rydberg interaction radius of nine lattice sites $R_c/a=9$ calculated via~\cite{Vaillant_2012}. The particular parameters used here were selected to be experimentally realistic and representative. 

\begin{table}[b]
\caption{\label{tab:table1}%
}
\begin{ruledtabular}
\begin{tabular}{lllllll}
\textrm{$\Omega_r/2\pi$}&
\textrm{$M$\footnote[1]{$M$: number of lattice sites, with $N\le M$ the number of atoms in the lattice}}&
\textrm{$a$\footnote[2]{$a$: Lattice constant fixed by "magic wavelength" $\lambda=\SI{813}{\nano\meter}$}}&
\textrm{$\Delta/2\pi$}&
$\textrm{P}_{\rm laser}$&
\textrm{$\omega_0$\footnote[3]{$\omega_0$: Rydberg laser beam waist}}&
\textrm{$n$\footnote[4]{$n$: Rydberg state principal quantum number}}\\
\colrule
$\SI{1.6}{\mega\hertz}$ & $10^3$ & $\SI{813/2}{\nano\meter}$  & $\SI{-16}{\mega\hertz}$ & $\SI{300}{\milli\watt}$ & $\SI{1}{\milli\meter}$ & 55\\
\end{tabular}
\end{ruledtabular}
\end{table}

As can be expected, the average minimal squeezing parameter $\langle \xi^2_{\textrm{min}}\rangle$ scales inversely with the filling probability $P_{\rm filling}$. Most of the observed variance in squeezing is due to the randomness in the number of atoms loaded into the lattice at a certain filling probability, as can be seen in Fig.~\ref{fig:xi2_vs_Pfilling}(b) where we plot the achieved squeezing as a function of the number of atoms randomly loaded into the lattice for each particular simulated configuration that contributed to the averaged values shown in Fig.~\ref{fig:xi2_vs_Pfilling}(a).

The dashed lines in Fig.~\ref{fig:xi2_vs_Pfilling}(b) represent limiting cases that add perspective to the random fractional filling simulation results.
The gold dashed line is the limit where all atoms in the lattice interact with all others with equal interaction strength $V_0 = \frac{\hbar\Omega_r^4}{8\Delta^3}$ and represents the theoretical best possible squeezing with this one-axis twisting approach. For the minimal squeezing $\xi^2_{\textrm{min}}$ (Eq.~\eqref{eq:xi}), the analytical solution reduces to:

\begin{subequations}
\label{eq:Rcinf}
\begin{eqnarray}
	&&\langle\hat{J}_z\rangle = \frac{-N}{2}\cos^{N-1}\left(\frac{V_0\tau}{2\hbar}\right), \\
	&&\langle\hat{J}_x^2\rangle = \frac{N}{4} + \frac{N(N-1)}{8}\left(1-\cos^{N-2}\left(2\frac{V_0\tau}{2\hbar}\right)\right), \\
	&&\langle\hat{J}_y^2\rangle = \frac{N}{4} , \\
	&&\langle\hat{J}_x\hat{J}_y + \hat{J}_y\hat{J}_x\rangle = \frac{-N(N-1)}{2}\sin\left(\frac{V_0\tau}{2\hbar}\right)\cos^{N-2}\left(\frac{V_0\tau}{2\hbar}\right) .\nonumber \\
\end{eqnarray}
\end{subequations}
    
\noindent In this limit the interaction potential has an infinite radius and the squeezing is not influenced by the geometry of the lattice nor by random filling.

In the experimentally relevant case where we are limited by finite-range Rydberg interactions, the optimal possible random fractionally-filled configuration to maximize the degree of squeezing is to place all the atoms next to each other without gaps, as is possible in a tweezer array where the traps can be rearranged to eliminate empty sites. This lower bound corresponds to the purple dashed line in Fig.~\ref{fig:xi2_vs_Pfilling}(b).

The inverse scaling of the minimal squeezing parameter $\xi^2_{\rm min}$ with filling fraction can mostly be explained by the increase in average distance between nearest neighbor atoms.
The average spacing between atoms in a random fractionally filled lattice is $\langle g \rangle= \frac{a}{P_{\textrm{filling}}} = \frac{M a}{\langle N \rangle}$ (with $\langle N \rangle$ the average number of atoms loaded into the lattice at filling probability $P_{\textrm{filling}}$). It is interesting to compare the random fractionally filled lattice with a fully filled lattice with rescaled lattice constant $a'=\langle g \rangle$. As illustrated by the solid green line in Fig~\ref{fig:xi2_vs_Pfilling}(b), most of the effect of random fractional filling on the spin squeezing can be explained by the average gap in between atoms reducing the effective interaction range $R_c/a$.

With this rescaled lattice model it is also possible to employ the computationally simpler Heaviside approximation to estimate the effects of random fractional lattice filling on the spin squeezing. The step-like behavior stems from the instant interaction cut-off beyond the Rydberg interaction radius $R_c$. In large fully filled one-dimensional lattice, adding more atoms barely improves the sqeezing as illustrated by the lower bound limit in Fig.~\ref{fig:xi2_vs_Pfilling}(b). This explains why the Heaviside approximation remains approximately constant until the rescaled lattice gap reduces enough to add an extra atom inside the interaction radius.

The rescaled lattice model captures most of the increase of the squeezing parameter in partially filled lattices, and it does allow for fast estimates through the Heaviside approximation. However it does systematically underestimate the increase, as shown by the green line in Fig.~\ref{fig:xi2_vs_Pfilling}(b).  
In appendix~\ref{app:fitting} we present better estimates in the form of empirical fitting functions for the average squeezing $\langle \xi^2_{\rm min}\rangle$ Eq.~\eqref{eq:xi2_fit} and squeezing time $\tau$ Eq.~\eqref{eq:tau_fit} in randomly filled one-dimensional lattices with various sizes $M \ge 400$ and interaction radii $R_c/a \in [1;30]$ as a function of the filling fraction $x=N/M$. 

\subsection{Squeezing Time and Rydberg State Decay}
In the far off-resonance Rydberg dressing regime, we ignored decay from the finite lifetime Rydberg state $\ket{r}$. A figure of merit for this assumption is the ratio $\tilde{\tau}/\tau_{\rm opt}$, where $\tilde{\tau} = \frac{4\Delta^2}{\Omega_r^2}\tau_{\ket{r}}$ is the enhanced Rydberg state lifetime and $\tau_{\ket{r}} \gtrsim \SI{23}{\micro \second}$ is the lifetime for the strontium Rydberg state with principal quantum number $n=55$. The simulated optimal squeezing time is $\tau_{\rm opt} \approx \SI{340}{\micro\second}$ for the partially filled one-dimensional lattice experiment considered, resulting in the figure of merit $\frac{\tilde{\tau}}{\tau_{\rm opt}} \approx 27.1 \gg 1$. 
In~\cite{PhysRevA.65.041803} the effects of incoherent decay from the Rydberg state on the squeezing are discussed in a simplified treatment. With the assumption that the number of atoms lost due to Rydberg decay equals $\frac{N}{\tilde{\tau}/\tau_{\rm opt}}$, we adopt their approximate estimate for the increase in squeezing parameter as 

\begin{equation}
    \Bar{\xi^{2}} = \xi^2 + \frac{1}{\tilde{\tau}/\tau_{\rm opt}}.
\label{eq:xi2_decay}
\end{equation}

When we apply this estimate to the simulation result of Fig.~\ref{fig:xi2_vs_Pfilling}(a) at $P_{\rm filling}=0.5$, we get $\langle \Bar{\xi^{2}}\rangle \approx 0.337$ which is an increase of $12\%$. 
The impact of incoherent Rydberg decay during spin squeezing can be reduced by employing states with higher principal quantum numbers $n$ which have longer life times, or by driving the Rydberg transition closer to resonance to improve the figure of merit $\frac{\tilde{\tau}}{\tau_{\rm opt}}$ as illustrated by Fig.~\ref{fig:detuning} in the appendix. We note that this simplistic treatment neglects the impact of accidental so-called ``spaghetti'' resonances between pairs of nearby atoms \cite{jau2016} which may increase the number of Rydberg atoms that are actually generated during dressing, as well as the dephasing that contaminant Rydberg states can induce \cite{goldschmidt2016,boulier2017}. We leave these important considerations for future investigation.

\subsection{\label{sec:par_opt}Randomness and Experimental Parameters}
\begin{figure}[ht]
    \centering
    \includegraphics[page=3,width=0.5\textwidth]{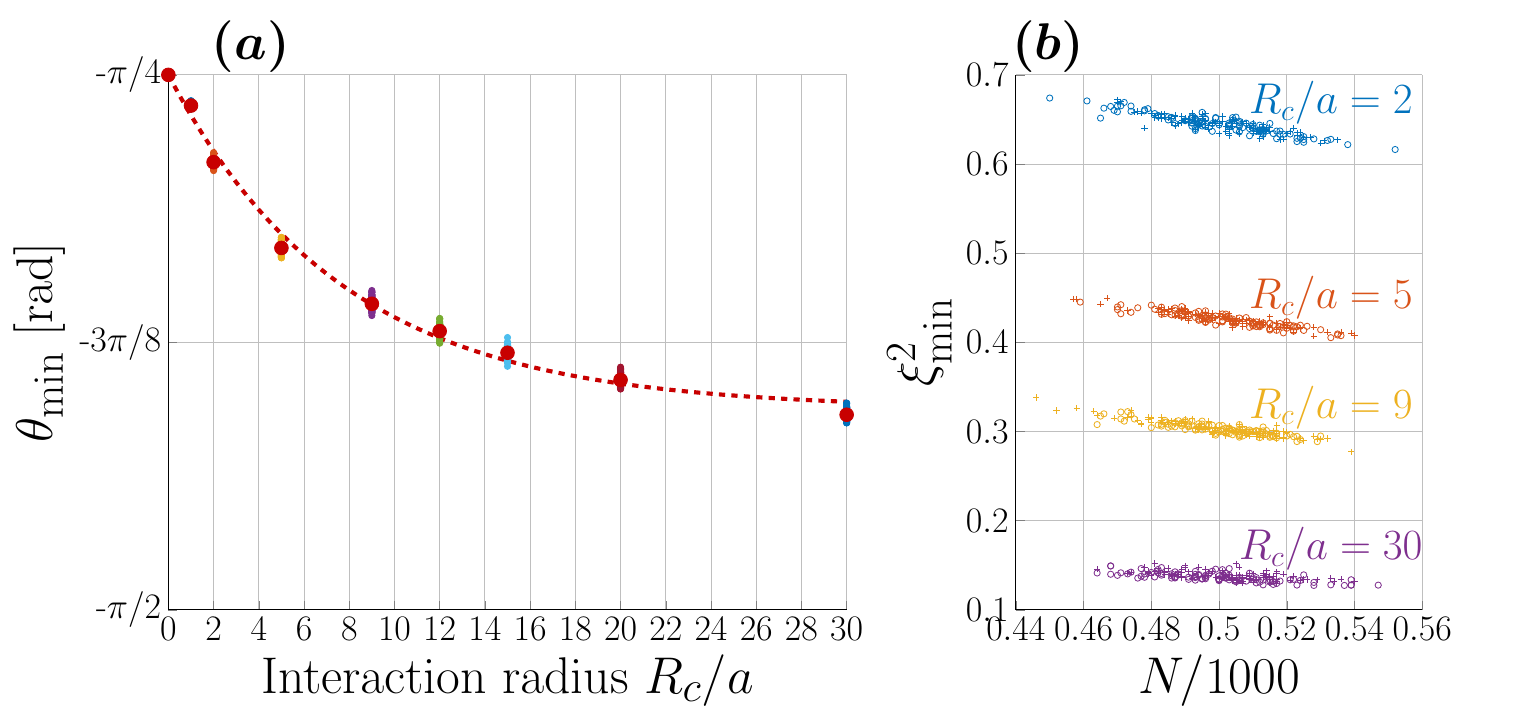}
    \caption{(a) The angle of minimal uncertainty $\theta_{\textrm{min}}$ as illustrated in Fig.~\ref{fig:intro} plotted for one-dimensional lattices of $M=1000$ sites, randomly filled with $P_{\textrm{filling}}=50\%$. The red line fits to the the average $\langle \theta_{\textrm{min}}\rangle$ of the random lattice configurations as a function of the Rydberg interaction radius $R_c/a$. (b) The squeezing $\xi^2$ obtained in these lattices, where the filled circle data points are the results of simulations with $\tau_{\textrm{opt}}$ and $\theta_{\textrm{min}}$ optimized for each individual random lattice configuration, while the cross data point simulations used a fixed $\tau$ and $\theta$ derived from the fitting functions for all the random lattice configurations. 
    }
    \label{fig:theta_fit}
\end{figure}

Unlike in tweezer clocks, most optical lattice clocks currently lack single site single-shot atom detection and atom readout is destructive, meaning that the experimental parameters $\tau$ and $\theta_{\textrm{min}}$ can not be optimized each time the lattice is loaded with another random configuration during clock operation. A fixed average value for $\tau$ and $\theta_{\textrm{min}}$ must therefore be used, which could potentially reduce the useful degree of squeezing. Fortunately, we find that for realistic parameters the impact of using a fixed squeezing time $\tau$ and fixed angle of minimal uncertainty $\theta_{\textrm{min}}$ for all random lattice configurations can be neglected, as illustrated by Fig.~\ref{fig:theta_fit}(b). The scatter points for optimized simulations overlap well with the cloud resulting from the well chosen fixed $\tau$ and $\theta_{\textrm{min}}$ simulations.\\

To determine the fixed $\tau$, we evaluate the general fitting function Eq.~\eqref{eq:tau_fit} at half filling $x=0.5$, 

\begin{equation}
\begin{split}
	\frac{V_0}{\hbar}\tau_{\textrm{opt, fit}}\left(\frac{R_c}{a}\right) &= 20.7\left(\frac{R_c}{a}\right)^{-1.49}
	\!\!\!
	\left(0.986 \cdot (1.016)^{R_c/a}-1\right)\\
	&+ 1.29\left(\frac{R_c}{a}\right)^{-0.635}.
\end{split}
	\label{eq:tau_fit_x05}
\end{equation}

\noindent We also present the fixed $\theta_{\textrm{min}}$ fitting function
\begin{equation}
\theta_{\textrm{min,fit}}\left(\frac{R_c}{a}\right) = 0.49\exp\left(-0.13\frac{R_c}{a}\right) - 0.49 - \frac{\pi}{4}.
\label{eq:theta_fit}
\end{equation}

\noindent This fitting function is illustrated as the dotted line in Fig.~\ref{fig:theta_fit}(a) going through the average optimal angle of minimal uncertainty for random lattice configurations.  This approximation is accurate 
for a wide range of interaction radii $R_c/a \lesssim 30$ for long one-dimensional lattices.

\section{Lattice Clock Stability Comparison}

As discussed above, the achievable degree of squeezing in partially filled lattices is lower than what can be achieved in their fully filled counterparts, which can potentially be constructed with optical tweezers~\cite{PhysRevX.9.041052,Endres1024}. However, the total number of tweezer traps that can be arranged into a single lattice is limited by the available laser power to about $N\approx 10^3$ atoms (current experimental demonstrations reach only hundreds of atoms~\cite{Barredo2018}), while lattice clocks can exceed those numbers, especially when two and three-dimensional lattices are considered \cite{campbell2017fermi,marti2018imaging, oelker2019demonstration}. 
Assuming zero-dead time, the clock fractional frequency stability $\sigma$ can be defined as a function of the averaging time $\tau_{\textrm{avg}}$,

\begin{equation}
	\sigma(\tau_{\textrm{avg}}) = \frac{\xi_{\textrm{min}}}{2\pi\nu_c\sqrt{N}\sqrt{\tau_{\textrm{avg}}T}}.
\label{eq:stability}
\end{equation}

\begin{figure}[ht]
    \centering
    \includegraphics[page=4,width=0.45\textwidth]{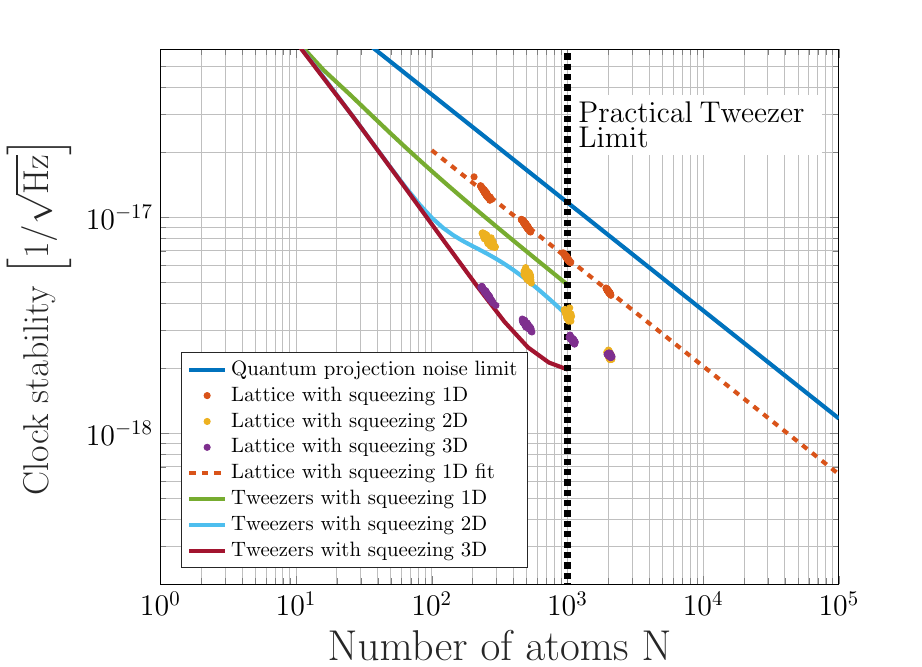}
    \caption{Simulation results of clock frequency stability as a function of the number of atoms in the lattice. This plot compares the performance of clocks with fully filled tweezer arrays or partially filled lattices ($P_{\textrm{filling}}=50\%$) of different dimensions (1D, 2D, 3D). The dashed fitting function is derived in appendix~\ref{app:fitting} and predicts the performance of lattice sizes that are no longer practical to fully simulate. The dotted black line illustrates the rough bound on the number of tweezer traps imposed by current laser technologies.
    }
    \label{fig:stability}
\end{figure}

\noindent In Fig.~\ref{fig:stability} $\sigma$ is plotted in units of $1/\sqrt{\SI{}{\hertz}}$  for fully and fractionally filled strontium optical  clocks with clock transition frequency $\nu_c = c/\lambda_c\approx\SI{429.228}{\tera\hertz}$ and interrogation time $T=\SI{1}{\second}$. For the purpose of a fair comparison we assume that the tweezer approach is also able to reach a Rydberg interaction radius of nine lattice sites $R_c/a=9$, although this requires stronger Rydberg interactions due to the strictly lower lattice constant achievable with tweezers at the clock magic wavelength. As reported in~\cite{mainpaper}, higher dimensional lattices achieve stronger squeezing, and result in better frequency stability. Partially filled one-dimensional lattices with $P_{\textrm{filling}}=0.5$ outperform the one-dimensional tweezers for atom numbers $N\ge2\cdot10^3$ (corresponding lattice size $M\ge4\cdot10^3$). The two- and three-dimensional tweezer arrays experience a temporary plateau in the spin squeezing when the lattice size is comparable to the number of atoms inside the Rydberg interaction radius $R_c$~\cite{mainpaper}. In this plateau region, the simulations show that the two-dimensional partially filled lattice performance becomes comparable to the three-dimensional lattice and could even beat the fully filled two-dimensional tweezer array. We anticipate the three-dimensional lattice to start outperforming the two-dimensional lattice again for larger atom numbers (that are computationally inconvenient to simulate). Figure~\ref{fig:stability} shows us that the choice of best lattice geometry is not trivial for fractionally filled lattices and is dependent on the system size.

\section{Conclusion}
Spin squeezing via Rydberg dressing is a convenient method with prospects for achieving large degrees of squeezing in state-of-the-art optical lattice clocks. We find that even in the presence of disorder due to random filling, partially filled lattices with larger atom numbers can achieve comparable or superior squeezed clock stabilities compared to the smaller fully filled arrays that can be realized with optical tweezers.
The spin squeezing is robust against small deviations in the optimal squeezing time $\tau$ and angle of minimal uncertainty $\theta_{\textrm{min}}$, meaning that experiments with fixed $\tau$ and $\theta_{\textrm{min}}$ for all randomly loaded lattice configurations remain viable.
To aid in the practical realization of spin-squeezing via Rydberg-dressing, we have also developed useful approximate methods as tools for experimental implementations in the form of the Heaviside approximation for fully filled one-dimensional lattices, and empirical fitting functions for partially filled one-dimensional lattices. We leave for future exploration the achievable squeezing when the Rydberg laser is brought closer to resonance, and the development of more exotic squeezing sequences to increase the degree of squeezing achievable with finite range interactions.

\FloatBarrier

\begin{acknowledgments}
The authors thank Adam Kaufman and Jeff Thompson for enlightening discussions and helpful comments on the manuscript. This work was supported in part by the NIST Precision Measurement Grants program, the Northwestern University Center for Fundamental Physics and the John Templeton Foundation through a Fundamental Physics grant, the Army Research Office through grant number W911NF1910084, and a Packard Fellowship for Science and Engineering. M.G.V.~was supported by DoE BES Materials and Chemical Sciences Research for Quantum Information Science program award No. DE-SC0019449. J.~VD acknowledges the support of the Belgian American Educational Foundation (B.A.E.F.) Fellowship.  
\end{acknowledgments}


\appendix

\section{\label{app:detuning}Optimal Rydberg Detuning}

\begin{figure}[ht]
    \centering
    \includegraphics[page=6,width=0.45\textwidth]{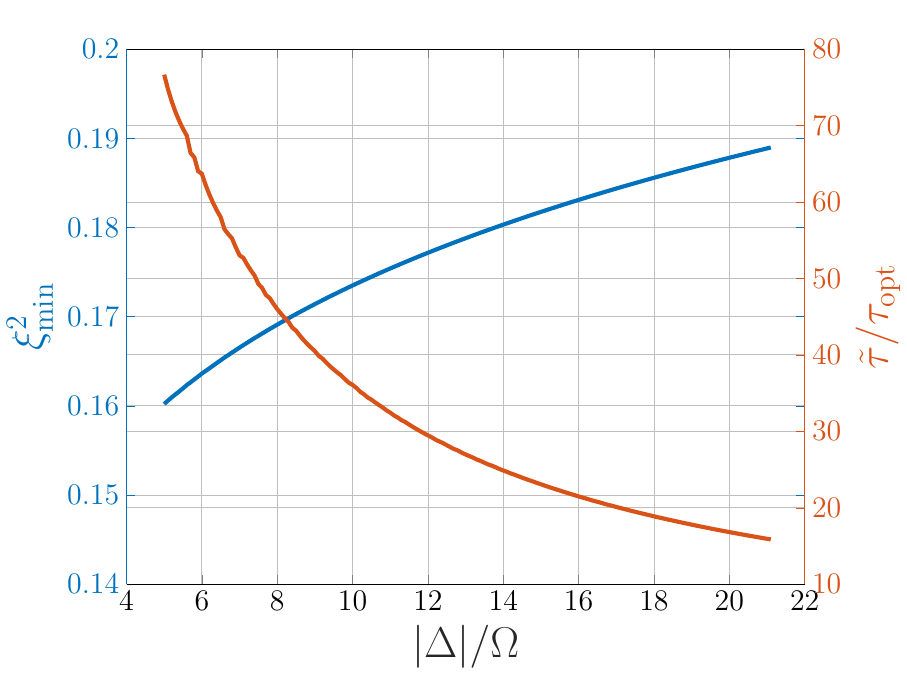}
    \caption{The squeezing parameter $\xi^2_{\textrm{min}}$ and the figure of merit $\tilde{\tau}/\tau_{\textrm{opt}}$ as a function of detuning $|\Delta|/\Omega_r$ from the Rydberg transition in a one-dimensional fully filled lattice of $M=1000$ sites with Rydberg interaction radius $R_c/a=9$.}
    \label{fig:detuning}
\end{figure}

The currently available theory developed to describe spin squeezing with Rydberg dressing requires far off-resonant drive of the Rydberg transition. In this regime the mixture of excited clock states and Rydberg states $\ket{\tilde{e}} \approx \ket{e} - \epsilon\ket{r}$ (with $\epsilon=\frac{\Omega_r}{2\Delta}$) has an improved lifetime $\tilde{\tau} = \frac{\tau_{\ket{r}}}{\epsilon^2} \approx \frac{\SI{23}{\micro\second}}{\epsilon^2}$. Which is argued to be necessary to facilitate long enough squeezing times. To find the optimal detuning of our Rydberg laser, we simulated the obtainable squeezing with the parameters from Table \ref{tab:table1} for a range of detunings $\Delta$.\\
Note that many influential parameters for spin squeezing depend on the detuning:

\begin{equation}
\begin{split}
V_0 &= \frac{\hbar\Omega_r^4}{8\Delta^3} \\
R_c &= \left|\frac{C_6}{2\hbar\Delta}\right|^{1/6} \\
\tilde{\tau} &= \frac{4\tau_{\ket{r}}\Delta^2}{\Omega_r^2},
\end{split}
\end{equation}

\noindent with $V_0$ the interaction potential inside the Rydberg interaction radius $R_c$ and $\tilde{\tau}$ the lifetime of the excited state mixture. Larger $V_0$ results in shorter optimal squeezing times, while larger interaction radius $R_c$ results in improved squeezing.\\
The result of the simulations is plotted in Fig.~\ref{fig:detuning}. We can conclude that squeezing $\xi^2_{\textrm{min}}$ improves at lower detuning, and surprisingly also the ratio life time to squeezing time $\tilde{\tau}/\tau_{\textrm{opt}}$ improves at lower detunings. This ratio $\tilde{\tau}/\tau_{\textrm{opt}}$ is a figure of merit for the assumption that decay from the Rydberg state $\ket{r}$ can be ignored during the squeezing operation.\\
The far off-resonance regime is not justifiable in terms of spin squeezing performance nor lifetime arguments, and new theory is required to describe the close to resonance regime.

\section{Rydberg Interaction Potential}

\subsection{\label{app:Vij_exact}Dipole-Dipole Interaction and the van der Waals Approximation}
The widely used Rydberg interaction potential of Eq.~\eqref{eq:Vij} is in fact an approximation. The interaction between Rydberg atoms is assumed to be dominated by the long range van der Waals approximation $\propto 1/r^6$, while the exact solution derived in~\cite{Saffman_2016} also includes the intermediate range dipole-dipole contribution $\propto 1/r^3$. We present this solution below in Eq.~\eqref{eq:Vij_exact}:

\begin{eqnarray}\label{eq:Vij_exact}
V_{ij}(r) &=& -\Delta + \frac{V_{dd}}{3} + \frac{2^{2/3}}{f}\left(\Delta^2-\Delta V_{dd} + V_{dd}^2/3 + |\Omega_r|^2\right) \nonumber\\
&+& \frac{2^{1/3}f}{6}.
\end{eqnarray}
with
\begin{eqnarray}
V_{dd} &=& \frac{\delta}{2}-\frac{\delta}{2}\sqrt{1+\left(\frac{x_c}{r}\right)^6} \nonumber\\
f &=& \bigg[18\Delta V_{dd}(\Delta-V_{dd})+4V_{dd}^3-9V_{dd}|\Omega_r|^2 \nonumber\\
&&+\big[V_{dd}^2(18\Delta^2-18\Delta V_{dd}+4V_{dd}^2-9|\Omega_r|^2)^2 \nonumber \nonumber\\
&&-16(3\Delta^2-3\Delta V_{dd} + V_{dd}^2 + 3|\Omega_r|^2)^3\big]^{1/2}\bigg]^{1/3}. \nonumber
\end{eqnarray}

\begin{figure}[ht]
    \centering
    \includegraphics[page=9,width=0.45\textwidth]{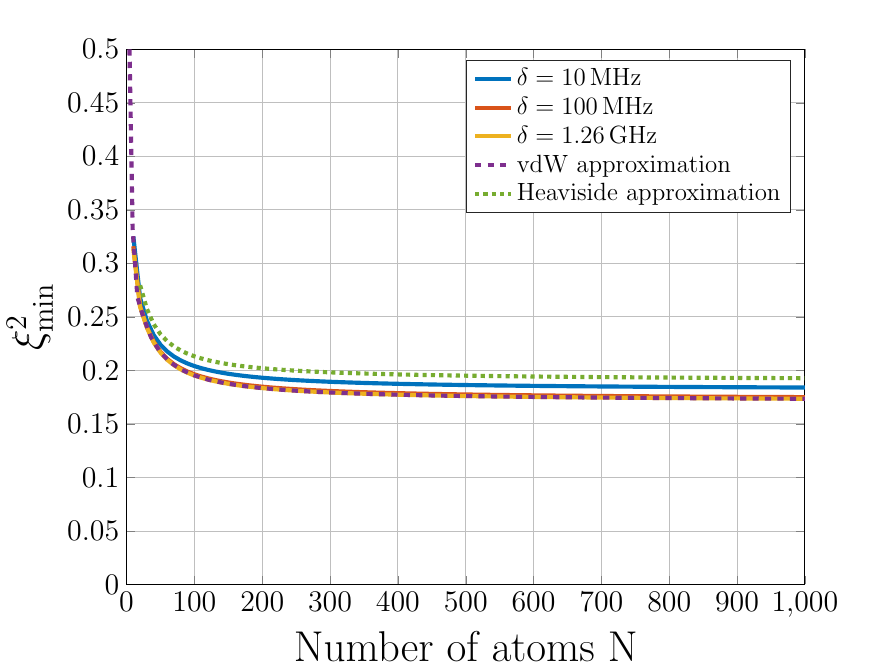}
    \caption{The squeezing parameter $\xi^2_{\rm min}$ in a one-dimensional fully filled lattice calculated with the solution Eq.~\eqref{eq:solution}. We plot the impact of using the exact Rydberg interaction potential $V_{ij}$ of Eq.~\eqref{eq:Vij_exact} with different F\"{o}rster defects $\delta \in \{\SI{10}{\mega\hertz}, \SI{100}{\mega\hertz}, \SI{1.26}{\giga\hertz}\}$ to the van der Waals approximated potential Eq.~\eqref{eq:Vij} and the Heaviside solution.}
    \label{fig:xi2_Vij_exact}
\end{figure}

\begin{figure}[ht]
    \centering
    \includegraphics[page=10,width=0.45\textwidth]{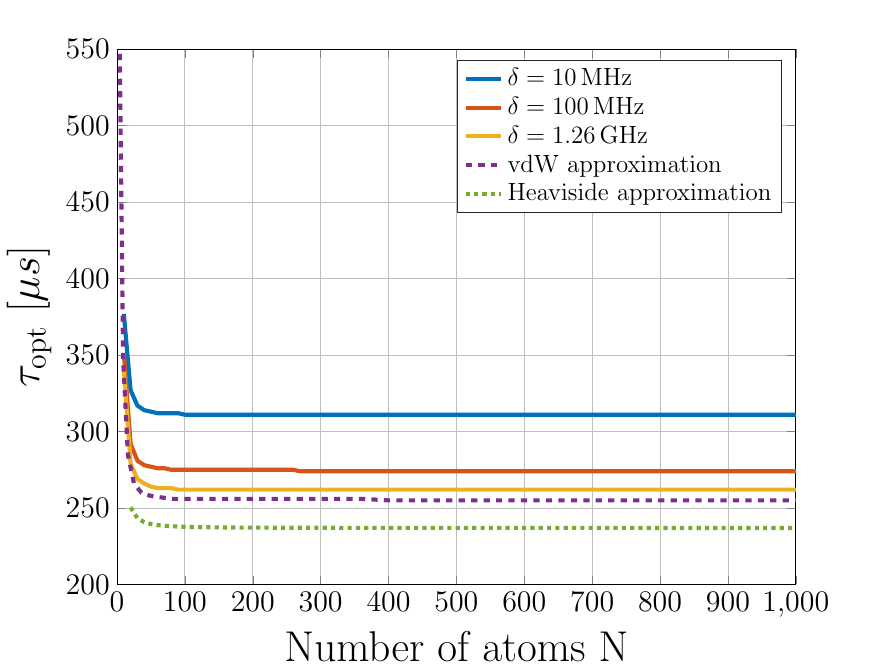}
    \caption{The optimal squeezing time $\tau_{\rm opt}$ in a one-dimensional fully filled lattice calculated with the solution Eq.~\eqref{eq:solution}. We plot the impact of using the exact Rydberg interaction potential $V_{ij}$ of Eq.~\eqref{eq:Vij_exact} with different F\"{o}rster defects $\delta \in \{\SI{10}{\mega\hertz}, \SI{100}{\mega\hertz}, \SI{1.26}{\giga\hertz}\}$ to the van der Waals approximated potential Eq.~\eqref{eq:Vij} and the Heaviside solution.}
    \label{fig:tau_Vij_exact}
\end{figure}

The Rydberg interaction potential $V_{ij}$ in Eq.~\eqref{eq:solution} arises from the position dependent energy shift of the excited clock state $\ket{e}$ due to the Rydberg dressing including dipole-dipole interactions of the dressed states. At long range, the dipole-dipole interaction is dominated by the van der Waals contribution $\propto 1/r^6$, and approximating the dipole-dipole interaction by just the van der Waals contribution is widely used, however this is not always justified. The simulations presented in Fig.~\ref{fig:xi2_Vij_exact} and Fig.~\ref{fig:tau_Vij_exact} show that the F\"{o}rster defect $\delta$ dictates the validity of the van der Waals approximation and $\delta$ should be calculated for the specific Rydberg state considered to justify the use of the van der Waals approximation.

The obtainable degree of squeezing is similar throughout the different interaction potentials, however the corresponding optimal squeezing time $\tau_{\rm opt}$ is strongly dependent on $\delta$ in the exact solution.

 \begin{figure}[ht]
    \centering
    \includegraphics[width=0.45\textwidth]{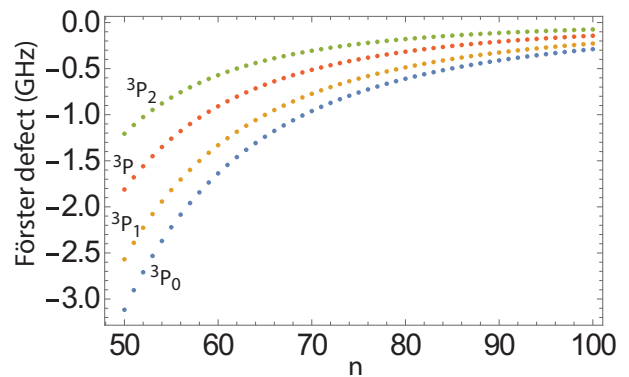}
    \caption{F\"orster defect of the $^3S_1$ series in the bosonic isotopes of Sr with nuclear spin zero. Calculations assume coupling to $^3P_{0,1,2}$ (blue, yellow, green dots) and a spin weighted average (red dots). }
    \label{fig:V(R)Mark}
\end{figure}

\subsection{\label{app:heaviside}Analytical Solution with Heaviside Approximation}

The analytical solution Eq.~\eqref{eq:solution} for spin squeezing via far off-resonance Rydberg dressing in atom lattices scales polynomially in the number of atoms. However for large lattices, this is still computationally very intensive. The interaction potential of the atoms can be approximated by a Heaviside function.

\begin{equation}
    V(r) \approx \begin{cases}\displaystyle 
            V_0 = \frac{\hbar\Omega_r^4}{8|\Delta|^3} &,~|r|\le R_c \\
            0 &,~\textrm{otherwise}\, .
    \end{cases}
    \label{eq:heaviside}
\end{equation}

\noindent With this interaction potential, it is possible to work out all the sums and products of the analytical solution Eq.~\eqref{eq:solution} for one-dimensional lattices and obtain a solution that can be evaluated instantly for all lattice sizes $M\ge2R_c/a$. Thanks to the binary interaction potential (two atoms either interact with strength $V_0$ or they don't interact) we can, without loss of generality, assume the interaction radius to be an integer times the lattice constant. \\
The analytical solution can be simplified using the identity
\begin{equation}
    \sum\limits_{k=1}^{K}\cos^k(\phi_0) = \frac{1-\cos^K(\phi_0)}{1-\cos(\phi_0)}
\end{equation}

By carefully addressing the edges of the lattice and making sure all atom interactions are accounted for, it is possible to evaluate all the summations and products of the analytical solution. The resulting expressions for the observables are presented below.

\begin{subequations}
\begin{widetext}
\begin{equation}
	\langle \hat{J}_z \rangle_h = \frac{-1}{2}\left[\left(N-2\frac{R_c}{a}\right)\cos^{2\frac{R_c}{a}}(\phi_0) + 2\cos^{\frac{R_c}{a}}(\phi_0)\frac{1-\cos^{\frac{R_c}{a}}(\phi_0)}{1-\cos(\phi_0)}\right],
	\label{eq:Jzh}
\end{equation}
\begin{equation}
	\langle \hat{J}_y^2\rangle_h = \frac{N}{4},
	\label{eq:Jy2h}
\end{equation}
\begin{equation}
\begin{split}
	\langle \hat{J}_x\hat{J}_y + \hat{J}_y\hat{J}_x \rangle_h = &-\left(N-3\frac{R_c}{a}\right)\frac{R_c}{a}\sin(\phi_0)\cos^{2\frac{R_c}{a}-1}(\phi_0) - \frac{R_c}{a}\sin(\phi_0)\cos^{\frac{R_c}{a}-1}(\phi_0)\frac{1-\cos^{\frac{R_c}{a}}(\phi_0)}{1-\cos(\phi_0)} \\
	&-\left(\frac{R_c}{a}\right)^2\sin(\phi_0)\cos^{2\frac{R_c}{a}-1}(\phi_0) - \left(\frac{R_c}{a}-1\right)\sin(\phi_0)\cos^{\frac{R_c}{a}}(\phi_0)\left(\frac{\cos^{\frac{R_c}{a}-1}-1}{\cos(\phi_0)-1}\right) \\
	&+\sin(\phi_0)\cos^{\frac{R_c}{a}}(\phi_0)\frac{\frac{R_c}{a}-2 +\cos^{\frac{R_c}{a}-1}(\phi_0) - \left(\frac{R_c}{a}-1\right)\cos(\phi_0)}{\left(\cos(\phi_0)-1\right)^2}\, ,
\end{split}
\label{eq:JxJyh}
\end{equation}
\begin{equation}
\begin{split}
	\langle \hat{J}_x^2 \rangle_h = &\frac{N}{4} + \frac{\left(N-4\frac{R_c}{a}\right)}{4}\frac{\cos^{2\frac{R_c}{a}}(\phi_0)}{\sin^2(\phi_0)}\left[1-\cos^{2\frac{R_c}{a}}(\phi_0) + \cos(2\phi_0)\left(\cos^{\frac{R_c}{a}}(2\phi_0) - \cos^{2\frac{R_c}{a}}(\phi_0)\right)\right] \\
	&+\frac{\left(N-3\frac{R_c}{a}\right)}{4}\frac{\cos^2(\phi_0)}{\sin^2(\phi_0)}\left[1-\cos^{2\frac{R_c}{a}}(\phi_0) + \cos^{\frac{R_c}{a}-1}(2\phi_0)\left(\cos^{\frac{R_c}{a}}(2\phi_0)-\cos^{2\frac{R_c}{a}}(\phi_0)\right)\right] \\
	&+\frac{1}{2}\Bigg[\left(\frac{R_c}{a}-\frac{\cos^{\frac{R_c}{a}-1}(2\phi_0)}{1-\cos(2\phi_0)}\right)\frac{\cos(\phi_0)\left(1-\cos^{\frac{R_c}{a}}(\phi_0)\right)}{1-\cos(\phi_0)} + \frac{\cos(\phi_0)\cos^{\frac{R_c}{a}-1}(2\phi_0)}{1-\cos(2\phi_0)}\left(\frac{\cos^{\frac{R_c}{a}}(2\phi_0)-\cos^{\frac{R_c}{a}}(\phi_0)}{\cos(2\phi_0)-\cos(\phi_0)}\right) \\
	&-\frac{\cos(\phi_0)\left(1-\cos^{\frac{R_c}{a}}(\phi_0)\right)-\frac{R_c}{a}\cos^{\frac{R_c}{a}+1}(\phi_0)(1-\cos(\phi_0))}{\left(1-\cos(\phi_0)\right)^2} \\
	&+ \frac{\cos(\phi_0)}{1-\cos(\phi_0)}\Bigg(\frac{\cos^{\frac{R_c}{a}}(\phi_0)-1}{\cos(\phi_0)-1} + \frac{\cos(\phi_0)}{\sin^2(\phi_0)}\left(\cos^{2\frac{R_c}{a}}(\phi_0)-1\right) \\
	&-\cos^{\frac{R_c}{a}-1}(2\phi_0)\left(\frac{\cos^{\frac{R_c}{a}}(\phi_0)-\cos^{\frac{R_c}{a}}(2\phi_0)}{\cos(\phi_0)-\cos(2\phi_0)}\right) + \frac{\cos(\phi_0)\cos^{\frac{R_c}{a}-1}(2\phi_0)}{\sin^{2}(\phi_0)}\left(\cos^{2\frac{R_c}{a}}(\phi_0)-\cos^{\frac{R_c}{a}}(2\phi_0)\right)\Bigg)\Bigg] \\
	&+\frac{1}{2}\left[\frac{\cos^{\frac{R_c}{a}}(\phi_0)\left(1-\cos^{\frac{R_c}{a}}(\phi_0)\right)}{\sin^2(\phi_0)(1-\cos(\phi_0))}\left(1-\cos^{2\frac{R_c}{a}}(\phi_0)+\cos(2\phi_0)\left(\cos^{\frac{R_c}{a}}(2\phi_0)-\cos^{2\frac{R_c}{a}}(\phi_0)\right)\right)\right] \\
	&+\frac{1+\cos^{2\frac{R_c}{a}-2}(\phi_0)}{4\tan^4(\phi_0)}\left[\frac{\cos^{2\frac{R_c}{a}}(\phi_0)-1}{\cos^2(\phi_0)}+\frac{R_c}{a}\tan^2(\phi_0)\right] \\
	&-\frac{\cos^{2\frac{R_c}{a}-2}(\phi_0)\cos(2\phi_0) + \cos^{\frac{R_c}{a}-1}(2\phi_0)}{4\tan^4(\phi_0)}\left[\frac{\cos(2\phi_0)}{\cos^2(\phi_0)}\left(\cos^{2\frac{R_c}{a}}(\phi_0)-\cos^{\frac{R_c}{a}}(2\phi_0)\right)-\frac{R_c}{a}\cos^{\frac{R_c}{a}}(2\phi_0)\tan^2(\phi_0)\right].
\end{split}
\end{equation}

\end{widetext}
\end{subequations}

\begin{figure}[ht]
    \centering
    \includegraphics[page=5,width=0.45\textwidth]{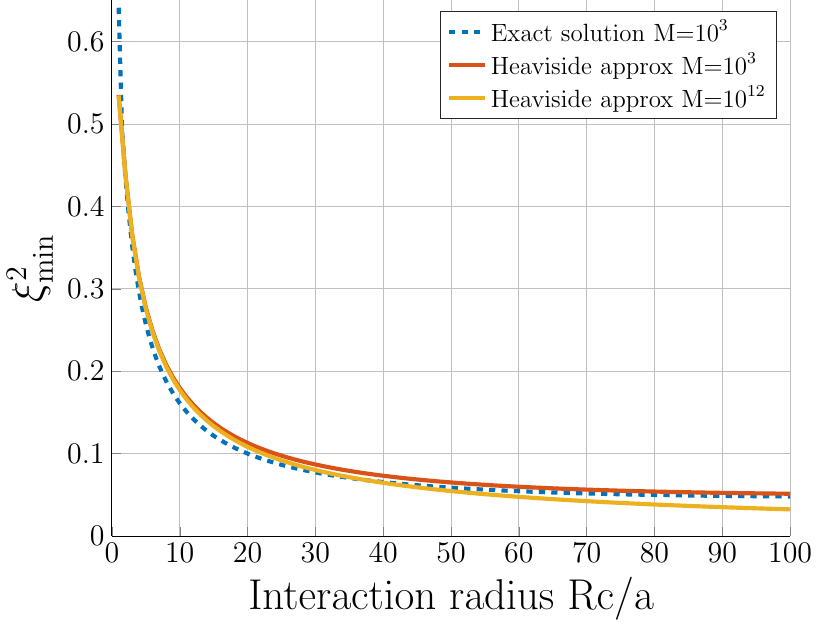}
    \caption{The squeezing parameter $\xi^2_{\textrm{min}}$ as a function of the interaction radius $R_c/a$ in a fully filled one-dimensional lattice of size $M$.}
    \label{fig:heaviside}
\end{figure}

Figure~\ref{fig:heaviside} compares the squeezing parameter calculated with the exact solution to the Heaviside approximation in a fully filled one-dimensional lattice of $M=10^3$ atoms. To showcase the power of this approximation, we plotted the same Heaviside solution with $M=10^{12}$ which would be impossible computationally with the exact solution, and is instantly evaluated with the approximation. We conclude that the Heaviside interaction potential is a reasonable approximation with a huge computational advantage. It is worth pointing out that the squeezing becomes almost invariant to the lattice size for large $M\gg 2R_c/a$ one-dimensional lattices.

\section{\label{app:fitting}Empirical Fitting Functions for Random Partial Filling of One-dimensional Lattices}

Squeezing is influenced by the statistics of the random lattice configurations. In the absence of a physical model describing these statistics properly, we can derive empirical fitting functions to describe the squeezing $\xi^2_{\textrm{min}}$ and the optimal squeezing time $\tau_{\textrm{opt}}$ for randomly filled one-dimensional lattices with various sizes and interaction radii $R_c/a$. These fitting functions are a useful tool to predict squeezing in randomly filled one-dimensional lattices without the need for simulations on cluster computers. We refer to the number of lattice sites as $M$, the number of atoms inside the lattice as $N$, with $N\le M$ and the filling fraction $x={N/M}$.
\\

We anticipate the squeezing parameter to be a function of the lattice size, the filling fraction and the interaction radius: $\xi^2_{\textrm{min}}(R_c/a,M,x)$. The fitting function has the general form of Eq.~\eqref{eq:xi2_fit} where the dependence on $R_c/a$ and $M$ is hidden in the fitting parameters $\{\alpha,\beta,\gamma,\lambda\}$:
\begin{equation}
	\xi^2_{\textrm{min}}(R_c/a,M,x) = \alpha \exp(-\beta x) + \gamma \exp(-\lambda x) .
	\label{eq:xi2_fit}
\end{equation}

From the requirement $\lim_{x \to 0}\xi^2_{\textrm{min}} = 1$ we can eliminate one degree of freedom in Eq.~\eqref{eq:xi2_fit}: $\gamma = 1-\alpha$. From simulations at full filling $x=1$ we conclude that the squeezing is approximately independent of $M$ in the case of large lattices ($M\gtrsim400$), this breaks down for very large $\frac{R_c}{a}$ (comparable to the system size), which we can ignore for the type of experiments we are considering. This means we can remove the dependence on $M$ in our fit and impose the dependence on $R_c/a$ at full filling with 
$$
\xi^2_{\textrm{min}}(\frac{R_c}{a},x=1) =  
0.6571\left(R_c/a\right)^{-0.656} + 0.01197
$$ 
by fitting to Fig. \ref{fig:heaviside} (note that we could have worked out an expression based on the Heaviside solution, but this would have been too bulky). We express the remaining fitting parameters $\{\beta, \lambda\}$ as functions of $R_c/a$ by fitting to multiple random filling simulations at different interaction radii. The resulting fitting parameter values are listed in Eq.~\eqref{eq:xi2fit_parameters}.
\begin{subequations}
\label{eq:xi2fit_parameters}
\begin{eqnarray}
\alpha &=&  \frac{0.657\left(\frac{R_c}{a}\right)^{-0.656} + 0.01197 - \exp(-\lambda)}{\exp(-\beta)-\exp(-\lambda)}\, ,  \\
	\beta &=& 0.293\frac{R_c}{a} + 5.297 \, ,\\
	\gamma &=& 1 - \alpha , \\
	\lambda &=& 1.14 - 2\exp\left(-0.89\frac{R_c}{a}\right).
\end{eqnarray}
\end{subequations}

\begin{figure}[ht]
    \centering
    \includegraphics[page=7,width=0.45\textwidth]{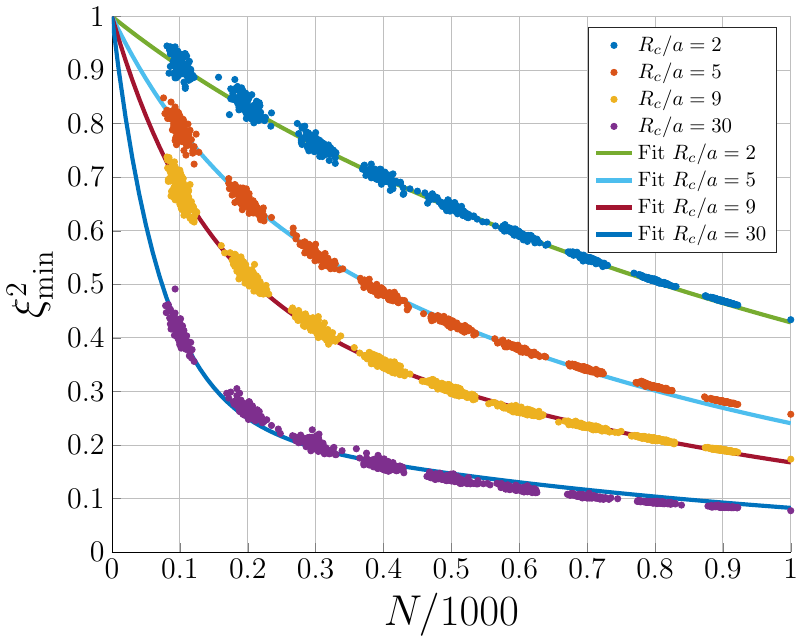}
    \caption{The squeezing parameter $\xi^2_{\textrm{min}}$ in random fractionally filled one-dimensional lattices of size $M=10^3$ simulated for different Rydberg interaction radii $R_c$ together with the corresponding fitting function from Eq.~\eqref{eq:xi2_fit}.}
    \label{fig:xi2_fit}
\end{figure}

Figure \ref{fig:xi2_fit} illustrates the fitting function of Eq.~\eqref{eq:xi2_fit} for different interaction radii ${R_c/a} \in \{2,5,9,30\}$. The fits perform well and can be evaluated instantly to predict the average squeezing obtainable in randomly filled one-dimensional lattices of various sizes $M \ge 400$ and with various interaction radii $\frac{R_c}{a} \in [1;30]$, instead of performing hundreds/thousands of simulations taking several hours each. 
Similarly we present a fitting function for the corresponding optimal squeezing times Eq.~\eqref{eq:tau_fit}.
\begin{equation}
\begin{split}
&\frac{V_0}{\hbar}\tau_{\textrm{opt}}\left(\frac{R_c}{a},M,x\right) =  \mu x^{-\nu} + 1.29\left(\frac{R_c}{a}\right)^{-0.635} - \mu \, , 
\\
&\mu =  20.7\left(\frac{R_c}{a}\right)^{-1.49}, \quad
\nu = 0.0229\frac{R_c}{a} - 0.0198.
\label{eq:tau_fit}
\end{split}
\end{equation}

\begin{figure}[ht]
    \centering
    \includegraphics[page=8,width=0.45\textwidth]{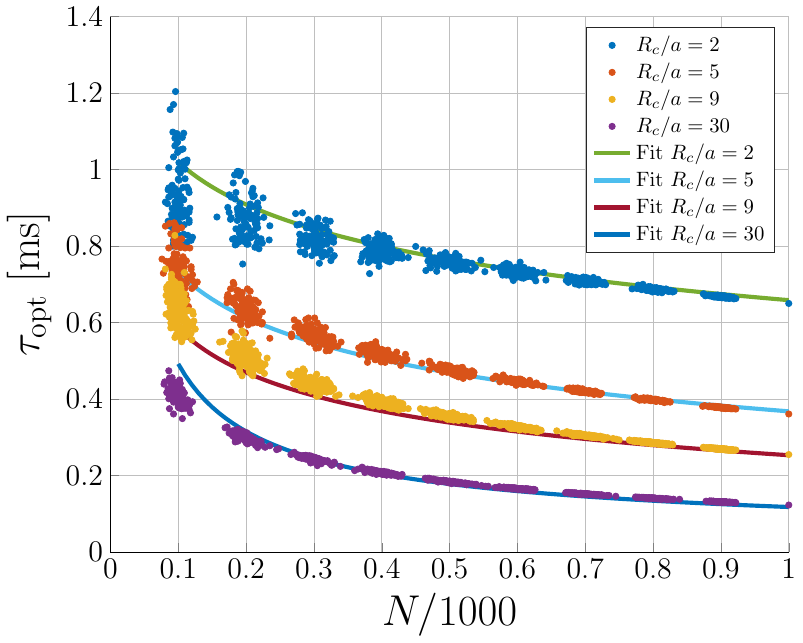}
    \caption{The optimal squeezing time $\tau_{\rm opt}$ in random fractionally filled one-dimensional lattices of size $M=10^3$ simulated for different Rydberg interaction radii $R_c$ together with the corresponding fitting function from Eq.~\eqref{eq:tau_fit}.}
    \label{fig:tau_fit}
\end{figure}

The performance of fitting function Eq.~\eqref{eq:tau_fit} is illustrated in Figure \ref{fig:tau_fit}, and performs well except for small filling fractions where the variance of optimal squeezing time $\tau_{\textrm{opt}}$ for randomly filled lattices becomes large.

\FloatBarrier

\bibliography{apssamp.bib}

\providecommand{\noopsort}[1]{}\providecommand{\singleletter}[1]{#1}%
\begin{thebibliography}{42}%
\makeatletter
\providecommand \@ifxundefined [1]{%
 \@ifx{#1\undefined}
}%
\providecommand \@ifnum [1]{%
 \ifnum #1\expandafter \@firstoftwo
 \else \expandafter \@secondoftwo
 \fi
}%
\providecommand \@ifx [1]{%
 \ifx #1\expandafter \@firstoftwo
 \else \expandafter \@secondoftwo
 \fi
}%
\providecommand \natexlab [1]{#1}%
\providecommand \enquote  [1]{``#1''}%
\providecommand \bibnamefont  [1]{#1}%
\providecommand \bibfnamefont [1]{#1}%
\providecommand \citenamefont [1]{#1}%
\providecommand \href@noop [0]{\@secondoftwo}%
\providecommand \href [0]{\begingroup \@sanitize@url \@href}%
\providecommand \@href[1]{\@@startlink{#1}\@@href}%
\providecommand \@@href[1]{\endgroup#1\@@endlink}%
\providecommand \@sanitize@url [0]{\catcode `\\12\catcode `\$12\catcode
  `\&12\catcode `\#12\catcode `\^12\catcode `\_12\catcode `\%12\relax}%
\providecommand \@@startlink[1]{}%
\providecommand \@@endlink[0]{}%
\providecommand \url  [0]{\begingroup\@sanitize@url \@url }%
\providecommand \@url [1]{\endgroup\@href {#1}{\urlprefix }}%
\providecommand \urlprefix  [0]{URL }%
\providecommand \Eprint [0]{\href }%
\providecommand \doibase [0]{https://doi.org/}%
\providecommand \selectlanguage [0]{\@gobble}%
\providecommand \bibinfo  [0]{\@secondoftwo}%
\providecommand \bibfield  [0]{\@secondoftwo}%
\providecommand \translation [1]{[#1]}%
\providecommand \BibitemOpen [0]{}%
\providecommand \bibitemStop [0]{}%
\providecommand \bibitemNoStop [0]{.\EOS\space}%
\providecommand \EOS [0]{\spacefactor3000\relax}%
\providecommand \BibitemShut  [1]{\csname bibitem#1\endcsname}%
\let\auto@bib@innerbib\@empty
\bibitem [{\citenamefont {Bloom}\ \emph {et~al.}(2014)\citenamefont {Bloom},
  \citenamefont {Nicholson}, \citenamefont {Williams}, \citenamefont
  {Campbell}, \citenamefont {Bishof}, \citenamefont {Zhang}, \citenamefont
  {Zhang}, \citenamefont {Bromley},\ and\ \citenamefont {Ye}}]{atom_clock}%
  \BibitemOpen
  \bibfield  {author} {\bibinfo {author} {\bibfnamefont {B.}~\bibnamefont
  {Bloom}}, \bibinfo {author} {\bibfnamefont {T.}~\bibnamefont {Nicholson}},
  \bibinfo {author} {\bibfnamefont {J.}~\bibnamefont {Williams}}, \bibinfo
  {author} {\bibfnamefont {S.}~\bibnamefont {Campbell}}, \bibinfo {author}
  {\bibfnamefont {M.}~\bibnamefont {Bishof}}, \bibinfo {author} {\bibfnamefont
  {X.}~\bibnamefont {Zhang}}, \bibinfo {author} {\bibfnamefont
  {W.}~\bibnamefont {Zhang}}, \bibinfo {author} {\bibfnamefont
  {S.}~\bibnamefont {Bromley}},\ and\ \bibinfo {author} {\bibfnamefont
  {J.}~\bibnamefont {Ye}},\ }\bibfield  {title} {\bibinfo {title} {An optical
  lattice clock with accuracy and stability at the $10^{-18}$ level},\ }\href
  {https://doi.org/10.1038/nature12941} {\bibfield  {journal} {\bibinfo
  {journal} {Nature}\ }\textbf {\bibinfo {volume} {506}},\ \bibinfo {pages}
  {71} (\bibinfo {year} {2014})}\BibitemShut {NoStop}%
\bibitem [{\citenamefont {Bothwell}\ \emph {et~al.}(2019)\citenamefont
  {Bothwell}, \citenamefont {Kedar}, \citenamefont {Oelker}, \citenamefont
  {Robinson}, \citenamefont {Bromley}, \citenamefont {Tew}, \citenamefont
  {Ye},\ and\ \citenamefont {Kennedy}}]{ab4089}%
  \BibitemOpen
  \bibfield  {author} {\bibinfo {author} {\bibfnamefont {T.}~\bibnamefont
  {Bothwell}}, \bibinfo {author} {\bibfnamefont {D.}~\bibnamefont {Kedar}},
  \bibinfo {author} {\bibfnamefont {E.}~\bibnamefont {Oelker}}, \bibinfo
  {author} {\bibfnamefont {J.}~\bibnamefont {Robinson}}, \bibinfo {author}
  {\bibfnamefont {S.}~\bibnamefont {Bromley}}, \bibinfo {author} {\bibfnamefont
  {W.}~\bibnamefont {Tew}}, \bibinfo {author} {\bibfnamefont {J.}~\bibnamefont
  {Ye}},\ and\ \bibinfo {author} {\bibfnamefont {C.}~\bibnamefont {Kennedy}},\
  }\bibfield  {title} {\bibinfo {title} {{JILA} {S}r{I} optical lattice clock
  with uncertainty of $2.0\times10^{-18}$},\ }\href
  {https://doi.org/10.1088/1681-7575/ab4089} {\bibfield  {journal} {\bibinfo
  {journal} {Metrologia}\ }\textbf {\bibinfo {volume} {56}},\ \bibinfo {pages}
  {065004} (\bibinfo {year} {2019})}\BibitemShut {NoStop}%
\bibitem [{\citenamefont {Marti}\ \emph {et~al.}(2018)\citenamefont {Marti},
  \citenamefont {Hutson}, \citenamefont {Goban}, \citenamefont {Campbell},
  \citenamefont {Poli},\ and\ \citenamefont {Ye}}]{marti2018imaging}%
  \BibitemOpen
  \bibfield  {author} {\bibinfo {author} {\bibfnamefont {G.~E.}\ \bibnamefont
  {Marti}}, \bibinfo {author} {\bibfnamefont {R.~B.}\ \bibnamefont {Hutson}},
  \bibinfo {author} {\bibfnamefont {A.}~\bibnamefont {Goban}}, \bibinfo
  {author} {\bibfnamefont {S.~L.}\ \bibnamefont {Campbell}}, \bibinfo {author}
  {\bibfnamefont {N.}~\bibnamefont {Poli}},\ and\ \bibinfo {author}
  {\bibfnamefont {J.}~\bibnamefont {Ye}},\ }\bibfield  {title} {\bibinfo
  {title} {Imaging optical frequencies with $100~\mu\textrm{Hz}$ precision and
  $1.1~\mu m$ resolution},\ }\href@noop {} {\bibfield  {journal} {\bibinfo
  {journal} {Phys. Rev. Let.}\ }\textbf {\bibinfo {volume} {120}},\ \bibinfo
  {pages} {103201} (\bibinfo {year} {2018})}\BibitemShut {NoStop}%
\bibitem [{\citenamefont {Oelker}\ \emph {et~al.}(2019)\citenamefont {Oelker},
  \citenamefont {Hutson}, \citenamefont {Kennedy}, \citenamefont {Sonderhouse},
  \citenamefont {Bothwell}, \citenamefont {Goban}, \citenamefont {Kedar},
  \citenamefont {Sanner}, \citenamefont {Robinson}, \citenamefont {Marti} \emph
  {et~al.}}]{oelker2019demonstration}%
  \BibitemOpen
  \bibfield  {author} {\bibinfo {author} {\bibfnamefont {E.}~\bibnamefont
  {Oelker}}, \bibinfo {author} {\bibfnamefont {R.}~\bibnamefont {Hutson}},
  \bibinfo {author} {\bibfnamefont {C.}~\bibnamefont {Kennedy}}, \bibinfo
  {author} {\bibfnamefont {L.}~\bibnamefont {Sonderhouse}}, \bibinfo {author}
  {\bibfnamefont {T.}~\bibnamefont {Bothwell}}, \bibinfo {author}
  {\bibfnamefont {A.}~\bibnamefont {Goban}}, \bibinfo {author} {\bibfnamefont
  {D.}~\bibnamefont {Kedar}}, \bibinfo {author} {\bibfnamefont
  {C.}~\bibnamefont {Sanner}}, \bibinfo {author} {\bibfnamefont
  {J.}~\bibnamefont {Robinson}}, \bibinfo {author} {\bibfnamefont
  {G.}~\bibnamefont {Marti}}, \emph {et~al.},\ }\bibfield  {title} {\bibinfo
  {title} {Demonstration of $4.8\times10^{-17}$ stability at 1 s for two
  independent optical clocks},\ }\href@noop {} {\bibfield  {journal} {\bibinfo
  {journal} {Nature Photonics}\ }\textbf {\bibinfo {volume} {13}},\ \bibinfo
  {pages} {714} (\bibinfo {year} {2019})}\BibitemShut {NoStop}%
\bibitem [{\citenamefont {Hume}\ and\ \citenamefont
  {Leibrandt}(2016)}]{hume2016probing}%
  \BibitemOpen
  \bibfield  {author} {\bibinfo {author} {\bibfnamefont {D.~B.}\ \bibnamefont
  {Hume}}\ and\ \bibinfo {author} {\bibfnamefont {D.~R.}\ \bibnamefont
  {Leibrandt}},\ }\bibfield  {title} {\bibinfo {title} {Probing beyond the
  laser coherence time in optical clock comparisons},\ }\href@noop {}
  {\bibfield  {journal} {\bibinfo  {journal} {Phys. Rev. A}\ }\textbf {\bibinfo
  {volume} {93}},\ \bibinfo {pages} {032138} (\bibinfo {year}
  {2016})}\BibitemShut {NoStop}%
\bibitem [{\citenamefont {Tan}\ \emph {et~al.}(2019)\citenamefont {Tan},
  \citenamefont {Kaewuam}, \citenamefont {Arnold}, \citenamefont {Chanu},
  \citenamefont {Zhang}, \citenamefont {Safronova},\ and\ \citenamefont
  {Barrett}}]{tan2019suppressing}%
  \BibitemOpen
  \bibfield  {author} {\bibinfo {author} {\bibfnamefont {T.~R.}\ \bibnamefont
  {Tan}}, \bibinfo {author} {\bibfnamefont {R.}~\bibnamefont {Kaewuam}},
  \bibinfo {author} {\bibfnamefont {K.~J.}\ \bibnamefont {Arnold}}, \bibinfo
  {author} {\bibfnamefont {S.~R.}\ \bibnamefont {Chanu}}, \bibinfo {author}
  {\bibfnamefont {Z.}~\bibnamefont {Zhang}}, \bibinfo {author} {\bibfnamefont
  {M.}~\bibnamefont {Safronova}},\ and\ \bibinfo {author} {\bibfnamefont
  {M.~D.}\ \bibnamefont {Barrett}},\ }\bibfield  {title} {\bibinfo {title}
  {Suppressing inhomogeneous broadening in a lutetium multi-ion optical
  clock},\ }\href@noop {} {\bibfield  {journal} {\bibinfo  {journal} {Phys.
  Rev. Let.}\ }\textbf {\bibinfo {volume} {123}},\ \bibinfo {pages} {063201}
  (\bibinfo {year} {2019})}\BibitemShut {NoStop}%
\bibitem [{\citenamefont {Young}\ \emph {et~al.}(2020)\citenamefont {Young},
  \citenamefont {Eckner}, \citenamefont {Milner}, \citenamefont {Kedar},
  \citenamefont {Norcia}, \citenamefont {Oelker}, \citenamefont {Schine},
  \citenamefont {Ye},\ and\ \citenamefont {Kaufman}}]{young2020tweezer}%
  \BibitemOpen
  \bibfield  {author} {\bibinfo {author} {\bibfnamefont {A.~W.}\ \bibnamefont
  {Young}}, \bibinfo {author} {\bibfnamefont {W.~J.}\ \bibnamefont {Eckner}},
  \bibinfo {author} {\bibfnamefont {W.~R.}\ \bibnamefont {Milner}}, \bibinfo
  {author} {\bibfnamefont {D.}~\bibnamefont {Kedar}}, \bibinfo {author}
  {\bibfnamefont {M.~A.}\ \bibnamefont {Norcia}}, \bibinfo {author}
  {\bibfnamefont {E.}~\bibnamefont {Oelker}}, \bibinfo {author} {\bibfnamefont
  {N.}~\bibnamefont {Schine}}, \bibinfo {author} {\bibfnamefont
  {J.}~\bibnamefont {Ye}},\ and\ \bibinfo {author} {\bibfnamefont {A.~M.}\
  \bibnamefont {Kaufman}},\ }\bibfield  {title} {\bibinfo {title} {A tweezer
  clock with half-minute atomic coherence at optical frequencies and high
  relative stability},\ }\href@noop {} {\bibfield  {journal} {\bibinfo
  {journal} {arXiv preprint arXiv:2004.06095}\ } (\bibinfo {year}
  {2020})}\BibitemShut {NoStop}%
\bibitem [{\citenamefont {Clements}\ \emph {et~al.}(2020)\citenamefont
  {Clements}, \citenamefont {Kim}, \citenamefont {Cui}, \citenamefont {Hankin},
  \citenamefont {Brewer}, \citenamefont {Valencia}, \citenamefont {Chen},
  \citenamefont {Chou}, \citenamefont {Leibrandt},\ and\ \citenamefont
  {Hume}}]{clements2020lifetime}%
  \BibitemOpen
  \bibfield  {author} {\bibinfo {author} {\bibfnamefont {E.~R.}\ \bibnamefont
  {Clements}}, \bibinfo {author} {\bibfnamefont {M.~E.}\ \bibnamefont {Kim}},
  \bibinfo {author} {\bibfnamefont {K.}~\bibnamefont {Cui}}, \bibinfo {author}
  {\bibfnamefont {A.~M.}\ \bibnamefont {Hankin}}, \bibinfo {author}
  {\bibfnamefont {S.~M.}\ \bibnamefont {Brewer}}, \bibinfo {author}
  {\bibfnamefont {J.}~\bibnamefont {Valencia}}, \bibinfo {author}
  {\bibfnamefont {J.-S.}\ \bibnamefont {Chen}}, \bibinfo {author}
  {\bibfnamefont {C.-W.}\ \bibnamefont {Chou}}, \bibinfo {author}
  {\bibfnamefont {D.~R.}\ \bibnamefont {Leibrandt}},\ and\ \bibinfo {author}
  {\bibfnamefont {D.~B.}\ \bibnamefont {Hume}},\ }\bibfield  {title} {\bibinfo
  {title} {Lifetime-limited interrogation of two independent
  ${}^{27}\textrm{Al}^{+}$ clocks using correlation spectroscopy},\ }\href@noop
  {} {\bibfield  {journal} {\bibinfo  {journal} {arXiv preprint
  arXiv:2007.02193}\ } (\bibinfo {year} {2020})}\BibitemShut {NoStop}%
\bibitem [{\citenamefont {Takamoto}\ \emph {et~al.}(2020)\citenamefont
  {Takamoto}, \citenamefont {Ushijima}, \citenamefont {Ohmae}, \citenamefont
  {Yahagi}, \citenamefont {Kokado}, \citenamefont {Shinkai},\ and\
  \citenamefont {Katori}}]{relativity}%
  \BibitemOpen
  \bibfield  {author} {\bibinfo {author} {\bibfnamefont {M.}~\bibnamefont
  {Takamoto}}, \bibinfo {author} {\bibfnamefont {I.}~\bibnamefont {Ushijima}},
  \bibinfo {author} {\bibfnamefont {N.}~\bibnamefont {Ohmae}}, \bibinfo
  {author} {\bibfnamefont {T.}~\bibnamefont {Yahagi}}, \bibinfo {author}
  {\bibfnamefont {K.}~\bibnamefont {Kokado}}, \bibinfo {author} {\bibfnamefont
  {H.-a.}\ \bibnamefont {Shinkai}},\ and\ \bibinfo {author} {\bibfnamefont
  {H.}~\bibnamefont {Katori}},\ }\bibfield  {title} {\bibinfo {title} {Test of
  general relativity by a pair of transportable optical lattice clocks},\
  }\href {https://doi.org/10.1038/s41566-020-0619-8} {\bibfield  {journal}
  {\bibinfo  {journal} {Nature Photonics}\ }\textbf {\bibinfo {volume} {14}},\
  \bibinfo {pages} {411} (\bibinfo {year} {2020})}\BibitemShut {NoStop}%
\bibitem [{\citenamefont {Counts}\ \emph {et~al.}(2020)\citenamefont {Counts},
  \citenamefont {Hur}, \citenamefont {Craik}, \citenamefont {Jeon},
  \citenamefont {Leung}, \citenamefont {Berengut}, \citenamefont {Geddes},
  \citenamefont {Kawasaki}, \citenamefont {Jhe},\ and\ \citenamefont
  {Vuleti{\'c}}}]{counts2020evidence}%
  \BibitemOpen
  \bibfield  {author} {\bibinfo {author} {\bibfnamefont {I.}~\bibnamefont
  {Counts}}, \bibinfo {author} {\bibfnamefont {J.}~\bibnamefont {Hur}},
  \bibinfo {author} {\bibfnamefont {D.~P.~A.}\ \bibnamefont {Craik}}, \bibinfo
  {author} {\bibfnamefont {H.}~\bibnamefont {Jeon}}, \bibinfo {author}
  {\bibfnamefont {C.}~\bibnamefont {Leung}}, \bibinfo {author} {\bibfnamefont
  {J.~C.}\ \bibnamefont {Berengut}}, \bibinfo {author} {\bibfnamefont
  {A.}~\bibnamefont {Geddes}}, \bibinfo {author} {\bibfnamefont
  {A.}~\bibnamefont {Kawasaki}}, \bibinfo {author} {\bibfnamefont
  {W.}~\bibnamefont {Jhe}},\ and\ \bibinfo {author} {\bibfnamefont
  {V.}~\bibnamefont {Vuleti{\'c}}},\ }\bibfield  {title} {\bibinfo {title}
  {Evidence for nonlinear isotope shift in {Y}b$^{+}$ search for new boson},\
  }\href@noop {} {\bibfield  {journal} {\bibinfo  {journal} {Phys. Rev. Let.}\
  }\textbf {\bibinfo {volume} {125}},\ \bibinfo {pages} {123002} (\bibinfo
  {year} {2020})}\BibitemShut {NoStop}%
\bibitem [{\citenamefont {Ludlow}\ \emph {et~al.}(2015)\citenamefont {Ludlow},
  \citenamefont {Boyd}, \citenamefont {Ye}, \citenamefont {Peik},\ and\
  \citenamefont {Schmidt}}]{RevModPhys.87.637}%
  \BibitemOpen
  \bibfield  {author} {\bibinfo {author} {\bibfnamefont {A.}~\bibnamefont
  {Ludlow}}, \bibinfo {author} {\bibfnamefont {M.}~\bibnamefont {Boyd}},
  \bibinfo {author} {\bibfnamefont {J.}~\bibnamefont {Ye}}, \bibinfo {author}
  {\bibfnamefont {E.}~\bibnamefont {Peik}},\ and\ \bibinfo {author}
  {\bibfnamefont {P.}~\bibnamefont {Schmidt}},\ }\bibfield  {title} {\bibinfo
  {title} {Optical atomic clocks},\ }\href
  {https://doi.org/10.1103/RevModPhys.87.637} {\bibfield  {journal} {\bibinfo
  {journal} {Reviews of Modern Physics}\ }\textbf {\bibinfo {volume} {87}},\
  \bibinfo {pages} {637} (\bibinfo {year} {2015})}\BibitemShut {NoStop}%
\bibitem [{\citenamefont {Takano}\ \emph {et~al.}(2016)\citenamefont {Takano},
  \citenamefont {Takamoto}, \citenamefont {Ushijima}, \citenamefont {Ohmae},
  \citenamefont {Akatsuka}, \citenamefont {Yamaguchi}, \citenamefont
  {Kuroishi}, \citenamefont {Munekane}, \citenamefont {Miyahara},\ and\
  \citenamefont {Katori}}]{geodesy}%
  \BibitemOpen
  \bibfield  {author} {\bibinfo {author} {\bibfnamefont {T.}~\bibnamefont
  {Takano}}, \bibinfo {author} {\bibfnamefont {M.}~\bibnamefont {Takamoto}},
  \bibinfo {author} {\bibfnamefont {I.}~\bibnamefont {Ushijima}}, \bibinfo
  {author} {\bibfnamefont {N.}~\bibnamefont {Ohmae}}, \bibinfo {author}
  {\bibfnamefont {T.}~\bibnamefont {Akatsuka}}, \bibinfo {author}
  {\bibfnamefont {A.}~\bibnamefont {Yamaguchi}}, \bibinfo {author}
  {\bibfnamefont {Y.}~\bibnamefont {Kuroishi}}, \bibinfo {author}
  {\bibfnamefont {H.}~\bibnamefont {Munekane}}, \bibinfo {author}
  {\bibfnamefont {B.}~\bibnamefont {Miyahara}},\ and\ \bibinfo {author}
  {\bibfnamefont {H.}~\bibnamefont {Katori}},\ }\bibfield  {title} {\bibinfo
  {title} {Geopotential measurements with synchronously linked optical lattice
  clocks},\ }\href {https://doi.org/10.1038/nphoton.2016.159} {\bibfield
  {journal} {\bibinfo  {journal} {Nature Photonics}\ }\textbf {\bibinfo
  {volume} {10}},\ \bibinfo {pages} {662} (\bibinfo {year} {2016})}\BibitemShut
  {NoStop}%
\bibitem [{\citenamefont {Huntemann}\ \emph {et~al.}(2014)\citenamefont
  {Huntemann}, \citenamefont {Lipphardt}, \citenamefont {Tamm}, \citenamefont
  {Gerginov}, \citenamefont {Weyers},\ and\ \citenamefont
  {Peik}}]{new_physics}%
  \BibitemOpen
  \bibfield  {author} {\bibinfo {author} {\bibfnamefont {N.}~\bibnamefont
  {Huntemann}}, \bibinfo {author} {\bibfnamefont {B.}~\bibnamefont
  {Lipphardt}}, \bibinfo {author} {\bibfnamefont {C.}~\bibnamefont {Tamm}},
  \bibinfo {author} {\bibfnamefont {V.}~\bibnamefont {Gerginov}}, \bibinfo
  {author} {\bibfnamefont {S.}~\bibnamefont {Weyers}},\ and\ \bibinfo {author}
  {\bibfnamefont {E.}~\bibnamefont {Peik}},\ }\bibfield  {title} {\bibinfo
  {title} {Improved limit on a temporal variation of ${m}_{p}/{m}_{e}$ from
  comparisons of {Y}$\textrm{b}^{+}$ and {C}s atomic clocks},\ }\href
  {https://doi.org/10.1103/PhysRevLett.113.210802} {\bibfield  {journal}
  {\bibinfo  {journal} {Phys. Rev. Lett.}\ }\textbf {\bibinfo {volume} {113}},\
  \bibinfo {pages} {210802} (\bibinfo {year} {2014})}\BibitemShut {NoStop}%
\bibitem [{\citenamefont {Kolkowitz}\ \emph {et~al.}(2016)\citenamefont
  {Kolkowitz}, \citenamefont {Pikovski}, \citenamefont {Langellier},
  \citenamefont {Lukin}, \citenamefont {Walsworth},\ and\ \citenamefont
  {Ye}}]{gravitationalwaves}%
  \BibitemOpen
  \bibfield  {author} {\bibinfo {author} {\bibfnamefont {S.}~\bibnamefont
  {Kolkowitz}}, \bibinfo {author} {\bibfnamefont {I.}~\bibnamefont {Pikovski}},
  \bibinfo {author} {\bibfnamefont {N.}~\bibnamefont {Langellier}}, \bibinfo
  {author} {\bibfnamefont {M.~D.}\ \bibnamefont {Lukin}}, \bibinfo {author}
  {\bibfnamefont {R.~L.}\ \bibnamefont {Walsworth}},\ and\ \bibinfo {author}
  {\bibfnamefont {J.}~\bibnamefont {Ye}},\ }\bibfield  {title} {\bibinfo
  {title} {Gravitational wave detection with optical lattice atomic clocks},\
  }\href {https://doi.org/10.1103/PhysRevD.94.124043} {\bibfield  {journal}
  {\bibinfo  {journal} {Phys. Rev. D}\ }\textbf {\bibinfo {volume} {94}},\
  \bibinfo {pages} {124043} (\bibinfo {year} {2016})}\BibitemShut {NoStop}%
\bibitem [{\citenamefont {Giovannetti}\ \emph {et~al.}(2004)\citenamefont
  {Giovannetti}, \citenamefont {Lloyd},\ and\ \citenamefont
  {Maccone}}]{science.1104149}%
  \BibitemOpen
  \bibfield  {author} {\bibinfo {author} {\bibfnamefont {V.}~\bibnamefont
  {Giovannetti}}, \bibinfo {author} {\bibfnamefont {S.}~\bibnamefont {Lloyd}},\
  and\ \bibinfo {author} {\bibfnamefont {L.}~\bibnamefont {Maccone}},\
  }\bibfield  {title} {\bibinfo {title} {Quantum-enhanced measurements: Beating
  the standard quantum limit},\ }\href
  {https://doi.org/10.1126/science.1104149} {\bibfield  {journal} {\bibinfo
  {journal} {Science (New York, N.Y.)}\ }\textbf {\bibinfo {volume} {306}},\
  \bibinfo {pages} {1330} (\bibinfo {year} {2004})}\BibitemShut {NoStop}%
\bibitem [{\citenamefont {Pezz\`e}\ \emph {et~al.}(2018)\citenamefont
  {Pezz\`e}, \citenamefont {Smerzi}, \citenamefont {Oberthaler}, \citenamefont
  {Schmied},\ and\ \citenamefont {Treutlein}}]{RevModPhys.90.035005}%
  \BibitemOpen
  \bibfield  {author} {\bibinfo {author} {\bibfnamefont {L.}~\bibnamefont
  {Pezz\`e}}, \bibinfo {author} {\bibfnamefont {A.}~\bibnamefont {Smerzi}},
  \bibinfo {author} {\bibfnamefont {M.~K.}\ \bibnamefont {Oberthaler}},
  \bibinfo {author} {\bibfnamefont {R.}~\bibnamefont {Schmied}},\ and\ \bibinfo
  {author} {\bibfnamefont {P.}~\bibnamefont {Treutlein}},\ }\bibfield  {title}
  {\bibinfo {title} {Quantum metrology with nonclassical states of atomic
  ensembles},\ }\href {https://doi.org/10.1103/RevModPhys.90.035005} {\bibfield
   {journal} {\bibinfo  {journal} {Rev. Mod. Phys.}\ }\textbf {\bibinfo
  {volume} {90}},\ \bibinfo {pages} {035005} (\bibinfo {year}
  {2018})}\BibitemShut {NoStop}%
\bibitem [{\citenamefont {Kitagawa}\ and\ \citenamefont
  {Ueda}(1993)}]{PhysRevA.47.5138}%
  \BibitemOpen
  \bibfield  {author} {\bibinfo {author} {\bibfnamefont {M.}~\bibnamefont
  {Kitagawa}}\ and\ \bibinfo {author} {\bibfnamefont {M.}~\bibnamefont
  {Ueda}},\ }\bibfield  {title} {\bibinfo {title} {Squeezed spin states},\
  }\href {https://doi.org/10.1103/PhysRevA.47.5138} {\bibfield  {journal}
  {\bibinfo  {journal} {Phys. Rev. A}\ }\textbf {\bibinfo {volume} {47}},\
  \bibinfo {pages} {5138} (\bibinfo {year} {1993})}\BibitemShut {NoStop}%
\bibitem [{\citenamefont {Engelsen}\ \emph {et~al.}(2016)\citenamefont
  {Engelsen}, \citenamefont {Hosten}, \citenamefont {Krishnakumar},\ and\
  \citenamefont {Kasevich}}]{cavitysqueezing}%
  \BibitemOpen
  \bibfield  {author} {\bibinfo {author} {\bibfnamefont {N.~J.}\ \bibnamefont
  {Engelsen}}, \bibinfo {author} {\bibfnamefont {O.}~\bibnamefont {Hosten}},
  \bibinfo {author} {\bibfnamefont {R.}~\bibnamefont {Krishnakumar}},\ and\
  \bibinfo {author} {\bibfnamefont {M.~A.}\ \bibnamefont {Kasevich}},\
  }\bibfield  {title} {\bibinfo {title} {Engineering spin-squeezed states for
  quantum-enhanced atom interferometry},\ }in\ \href
  {https://doi.org/10.1364/CLEO_QELS.2016.FM3C.1} {\emph {\bibinfo {booktitle}
  {Conference on Lasers and Electro-Optics}}}\ (\bibinfo  {publisher} {Optical
  Society of America},\ \bibinfo {year} {2016})\ p.\ \bibinfo {pages}
  {FM3C.1}\BibitemShut {NoStop}%
\bibitem [{\citenamefont {Pedrozo-Peñafiel}\ \emph {et~al.}(2020)\citenamefont
  {Pedrozo-Peñafiel}, \citenamefont {Colombo}, \citenamefont {Shu},
  \citenamefont {Adiyatullin}, \citenamefont {Li}, \citenamefont {Mendez},
  \citenamefont {Braverman}, \citenamefont {Kawasaki}, \citenamefont
  {Akamatsu}, \citenamefont {Xiao},\ and\ \citenamefont
  {Vuletić}}]{2006.07501}%
  \BibitemOpen
  \bibfield  {author} {\bibinfo {author} {\bibfnamefont {E.}~\bibnamefont
  {Pedrozo-Peñafiel}}, \bibinfo {author} {\bibfnamefont {S.}~\bibnamefont
  {Colombo}}, \bibinfo {author} {\bibfnamefont {C.}~\bibnamefont {Shu}},
  \bibinfo {author} {\bibfnamefont {A.~F.}\ \bibnamefont {Adiyatullin}},
  \bibinfo {author} {\bibfnamefont {Z.}~\bibnamefont {Li}}, \bibinfo {author}
  {\bibfnamefont {E.}~\bibnamefont {Mendez}}, \bibinfo {author} {\bibfnamefont
  {B.}~\bibnamefont {Braverman}}, \bibinfo {author} {\bibfnamefont
  {A.}~\bibnamefont {Kawasaki}}, \bibinfo {author} {\bibfnamefont
  {D.}~\bibnamefont {Akamatsu}}, \bibinfo {author} {\bibfnamefont
  {Y.}~\bibnamefont {Xiao}},\ and\ \bibinfo {author} {\bibfnamefont
  {V.}~\bibnamefont {Vuletić}},\ }\href@noop {} {\bibinfo {title}
  {Entanglement-enhanced optical atomic clock}} (\bibinfo {year} {2020}),\
  \Eprint {https://arxiv.org/abs/arXiv:2006.07501} {arXiv:2006.07501}
  \BibitemShut {NoStop}%
\bibitem [{\citenamefont {Gil}\ \emph {et~al.}(2014)\citenamefont {Gil},
  \citenamefont {Mukherjee}, \citenamefont {Bridge}, \citenamefont {Jones},\
  and\ \citenamefont {Pohl}}]{mainpaper}%
  \BibitemOpen
  \bibfield  {author} {\bibinfo {author} {\bibfnamefont {L.~I.~R.}\
  \bibnamefont {Gil}}, \bibinfo {author} {\bibfnamefont {R.}~\bibnamefont
  {Mukherjee}}, \bibinfo {author} {\bibfnamefont {E.~M.}\ \bibnamefont
  {Bridge}}, \bibinfo {author} {\bibfnamefont {M.~P.~A.}\ \bibnamefont
  {Jones}},\ and\ \bibinfo {author} {\bibfnamefont {T.}~\bibnamefont {Pohl}},\
  }\bibfield  {title} {\bibinfo {title} {Spin squeezing in a {R}ydberg lattice
  clock},\ }\href {https://doi.org/10.1103/PhysRevLett.112.103601} {\bibfield
  {journal} {\bibinfo  {journal} {Phys. Rev. Lett.}\ }\textbf {\bibinfo
  {volume} {112}},\ \bibinfo {pages} {103601} (\bibinfo {year}
  {2014})}\BibitemShut {NoStop}%
\bibitem [{\citenamefont {Borish}\ \emph {et~al.}(2020)\citenamefont {Borish},
  \citenamefont {Marković}, \citenamefont {Hines}, \citenamefont {Rajagopal},\
  and\ \citenamefont {Schleier-Smith}}]{PhysRevLett.124.063601}%
  \BibitemOpen
  \bibfield  {author} {\bibinfo {author} {\bibfnamefont {V.}~\bibnamefont
  {Borish}}, \bibinfo {author} {\bibfnamefont {O.}~\bibnamefont {Marković}},
  \bibinfo {author} {\bibfnamefont {J.}~\bibnamefont {Hines}}, \bibinfo
  {author} {\bibfnamefont {S.}~\bibnamefont {Rajagopal}},\ and\ \bibinfo
  {author} {\bibfnamefont {M.}~\bibnamefont {Schleier-Smith}},\ }\bibfield
  {title} {\bibinfo {title} {Transverse-field {I}sing dynamics in a
  {R}ydberg-dressed atomic gas},\ }\href
  {https://doi.org/10.1103/PhysRevLett.124.063601} {\bibfield  {journal}
  {\bibinfo  {journal} {Phys. Rev. Let.}\ }\textbf {\bibinfo {volume} {124}}
  (\bibinfo {year} {2020})}\BibitemShut {NoStop}%
\bibitem [{\citenamefont {Madjarov}\ \emph {et~al.}(2020)\citenamefont
  {Madjarov}, \citenamefont {Covey}, \citenamefont {Shaw}, \citenamefont
  {Choi}, \citenamefont {Kale}, \citenamefont {Cooper}, \citenamefont
  {Pichler}, \citenamefont {Schkolnik}, \citenamefont {Williams},\ and\
  \citenamefont {Endres}}]{madjarov2020high}%
  \BibitemOpen
  \bibfield  {author} {\bibinfo {author} {\bibfnamefont {I.~S.}\ \bibnamefont
  {Madjarov}}, \bibinfo {author} {\bibfnamefont {J.~P.}\ \bibnamefont {Covey}},
  \bibinfo {author} {\bibfnamefont {A.~L.}\ \bibnamefont {Shaw}}, \bibinfo
  {author} {\bibfnamefont {J.}~\bibnamefont {Choi}}, \bibinfo {author}
  {\bibfnamefont {A.}~\bibnamefont {Kale}}, \bibinfo {author} {\bibfnamefont
  {A.}~\bibnamefont {Cooper}}, \bibinfo {author} {\bibfnamefont
  {H.}~\bibnamefont {Pichler}}, \bibinfo {author} {\bibfnamefont
  {V.}~\bibnamefont {Schkolnik}}, \bibinfo {author} {\bibfnamefont {J.~R.}\
  \bibnamefont {Williams}},\ and\ \bibinfo {author} {\bibfnamefont
  {M.}~\bibnamefont {Endres}},\ }\bibfield  {title} {\bibinfo {title}
  {High-fidelity entanglement and detection of alkaline-earth {R}ydberg
  atoms},\ }\href@noop {} {\bibfield  {journal} {\bibinfo  {journal} {Nature
  Physics}\ ,\ \bibinfo {pages} {857}} (\bibinfo {year} {2020})}\BibitemShut
  {NoStop}%
\bibitem [{\citenamefont {Nicholson}\ \emph {et~al.}(2015)\citenamefont
  {Nicholson}, \citenamefont {Campbell}, \citenamefont {Hutson}, \citenamefont
  {Marti}, \citenamefont {Bloom}, \citenamefont {McNally}, \citenamefont
  {Zhang}, \citenamefont {Barrett}, \citenamefont {Safronova}, \citenamefont
  {Strouse} \emph {et~al.}}]{nicholson2015systematic}%
  \BibitemOpen
  \bibfield  {author} {\bibinfo {author} {\bibfnamefont {T.}~\bibnamefont
  {Nicholson}}, \bibinfo {author} {\bibfnamefont {S.}~\bibnamefont {Campbell}},
  \bibinfo {author} {\bibfnamefont {R.}~\bibnamefont {Hutson}}, \bibinfo
  {author} {\bibfnamefont {G.}~\bibnamefont {Marti}}, \bibinfo {author}
  {\bibfnamefont {B.}~\bibnamefont {Bloom}}, \bibinfo {author} {\bibfnamefont
  {R.}~\bibnamefont {McNally}}, \bibinfo {author} {\bibfnamefont
  {W.}~\bibnamefont {Zhang}}, \bibinfo {author} {\bibfnamefont
  {M.}~\bibnamefont {Barrett}}, \bibinfo {author} {\bibfnamefont
  {M.}~\bibnamefont {Safronova}}, \bibinfo {author} {\bibfnamefont
  {G.}~\bibnamefont {Strouse}}, \emph {et~al.},\ }\bibfield  {title} {\bibinfo
  {title} {Systematic evaluation of an atomic clock at 2$\times10^{-18}$ total
  uncertainty},\ }\href@noop {} {\bibfield  {journal} {\bibinfo  {journal}
  {Nature communications}\ }\textbf {\bibinfo {volume} {6}},\ \bibinfo {pages}
  {6896} (\bibinfo {year} {2015})}\BibitemShut {NoStop}%
\bibitem [{\citenamefont {Campbell}\ \emph {et~al.}(2017)\citenamefont
  {Campbell}, \citenamefont {Hutson}, \citenamefont {Marti}, \citenamefont
  {Goban}, \citenamefont {Oppong}, \citenamefont {McNally}, \citenamefont
  {Sonderhouse}, \citenamefont {Robinson}, \citenamefont {Zhang}, \citenamefont
  {Bloom} \emph {et~al.}}]{campbell2017fermi}%
  \BibitemOpen
  \bibfield  {author} {\bibinfo {author} {\bibfnamefont {S.~L.}\ \bibnamefont
  {Campbell}}, \bibinfo {author} {\bibfnamefont {R.}~\bibnamefont {Hutson}},
  \bibinfo {author} {\bibfnamefont {G.}~\bibnamefont {Marti}}, \bibinfo
  {author} {\bibfnamefont {A.}~\bibnamefont {Goban}}, \bibinfo {author}
  {\bibfnamefont {N.~D.}\ \bibnamefont {Oppong}}, \bibinfo {author}
  {\bibfnamefont {R.}~\bibnamefont {McNally}}, \bibinfo {author} {\bibfnamefont
  {L.}~\bibnamefont {Sonderhouse}}, \bibinfo {author} {\bibfnamefont
  {J.}~\bibnamefont {Robinson}}, \bibinfo {author} {\bibfnamefont
  {W.}~\bibnamefont {Zhang}}, \bibinfo {author} {\bibfnamefont
  {B.}~\bibnamefont {Bloom}}, \emph {et~al.},\ }\bibfield  {title} {\bibinfo
  {title} {A {F}ermi-degenerate three-dimensional optical lattice clock},\
  }\href@noop {} {\bibfield  {journal} {\bibinfo  {journal} {Science}\ }\textbf
  {\bibinfo {volume} {358}},\ \bibinfo {pages} {90} (\bibinfo {year}
  {2017})}\BibitemShut {NoStop}%
\bibitem [{\citenamefont {Madjarov}\ \emph {et~al.}(2019)\citenamefont
  {Madjarov}, \citenamefont {Cooper}, \citenamefont {Shaw}, \citenamefont
  {Covey}, \citenamefont {Schkolnik}, \citenamefont {Yoon}, \citenamefont
  {Williams},\ and\ \citenamefont {Endres}}]{PhysRevX.9.041052}%
  \BibitemOpen
  \bibfield  {author} {\bibinfo {author} {\bibfnamefont {I.~S.}\ \bibnamefont
  {Madjarov}}, \bibinfo {author} {\bibfnamefont {A.}~\bibnamefont {Cooper}},
  \bibinfo {author} {\bibfnamefont {A.~L.}\ \bibnamefont {Shaw}}, \bibinfo
  {author} {\bibfnamefont {J.~P.}\ \bibnamefont {Covey}}, \bibinfo {author}
  {\bibfnamefont {V.}~\bibnamefont {Schkolnik}}, \bibinfo {author}
  {\bibfnamefont {T.~H.}\ \bibnamefont {Yoon}}, \bibinfo {author}
  {\bibfnamefont {J.~R.}\ \bibnamefont {Williams}},\ and\ \bibinfo {author}
  {\bibfnamefont {M.}~\bibnamefont {Endres}},\ }\bibfield  {title} {\bibinfo
  {title} {An atomic-array optical clock with single-atom readout},\ }\href
  {https://doi.org/10.1103/PhysRevX.9.041052} {\bibfield  {journal} {\bibinfo
  {journal} {Phys. Rev. X}\ }\textbf {\bibinfo {volume} {9}},\ \bibinfo {pages}
  {041052} (\bibinfo {year} {2019})}\BibitemShut {NoStop}%
\bibitem [{\citenamefont {Endres}\ \emph {et~al.}(2016)\citenamefont {Endres},
  \citenamefont {Bernien}, \citenamefont {Keesling}, \citenamefont {Levine},
  \citenamefont {Anschuetz}, \citenamefont {Krajenbrink}, \citenamefont
  {Senko}, \citenamefont {Vuletic}, \citenamefont {Greiner},\ and\
  \citenamefont {Lukin}}]{Endres1024}%
  \BibitemOpen
  \bibfield  {author} {\bibinfo {author} {\bibfnamefont {M.}~\bibnamefont
  {Endres}}, \bibinfo {author} {\bibfnamefont {H.}~\bibnamefont {Bernien}},
  \bibinfo {author} {\bibfnamefont {A.}~\bibnamefont {Keesling}}, \bibinfo
  {author} {\bibfnamefont {H.}~\bibnamefont {Levine}}, \bibinfo {author}
  {\bibfnamefont {E.~R.}\ \bibnamefont {Anschuetz}}, \bibinfo {author}
  {\bibfnamefont {A.}~\bibnamefont {Krajenbrink}}, \bibinfo {author}
  {\bibfnamefont {C.}~\bibnamefont {Senko}}, \bibinfo {author} {\bibfnamefont
  {V.}~\bibnamefont {Vuletic}}, \bibinfo {author} {\bibfnamefont
  {M.}~\bibnamefont {Greiner}},\ and\ \bibinfo {author} {\bibfnamefont {M.~D.}\
  \bibnamefont {Lukin}},\ }\bibfield  {title} {\bibinfo {title} {Atom-by-atom
  assembly of defect-free one-dimensional cold atom arrays},\ }\href
  {https://doi.org/10.1126/science.aah3752} {\bibfield  {journal} {\bibinfo
  {journal} {Science}\ }\textbf {\bibinfo {volume} {354}},\ \bibinfo {pages}
  {1024} (\bibinfo {year} {2016})}\BibitemShut {NoStop}%
\bibitem [{\citenamefont {Kaubruegger}\ \emph {et~al.}(2019)\citenamefont
  {Kaubruegger}, \citenamefont {Silvi}, \citenamefont {Kokail}, \citenamefont
  {van Bijnen}, \citenamefont {Rey}, \citenamefont {Ye}, \citenamefont
  {Kaufman},\ and\ \citenamefont {Zoller}}]{kaubruegger2019}%
  \BibitemOpen
  \bibfield  {author} {\bibinfo {author} {\bibfnamefont {R.}~\bibnamefont
  {Kaubruegger}}, \bibinfo {author} {\bibfnamefont {P.}~\bibnamefont {Silvi}},
  \bibinfo {author} {\bibfnamefont {C.}~\bibnamefont {Kokail}}, \bibinfo
  {author} {\bibfnamefont {R.}~\bibnamefont {van Bijnen}}, \bibinfo {author}
  {\bibfnamefont {A.~M.}\ \bibnamefont {Rey}}, \bibinfo {author} {\bibfnamefont
  {J.}~\bibnamefont {Ye}}, \bibinfo {author} {\bibfnamefont {A.~M.}\
  \bibnamefont {Kaufman}},\ and\ \bibinfo {author} {\bibfnamefont
  {P.}~\bibnamefont {Zoller}},\ }\bibfield  {title} {\bibinfo {title}
  {Variational spin-squeezing algorithms on programmable quantum sensors},\
  }\href@noop {} {\bibfield  {journal} {\bibinfo  {journal} {Phys. Rev. Let.}\
  }\textbf {\bibinfo {volume} {123}},\ \bibinfo {pages} {260505} (\bibinfo
  {year} {2019})}\BibitemShut {NoStop}%
\bibitem [{\citenamefont {Johnson}\ and\ \citenamefont
  {Rolston}(2010)}]{PhysRevA.82.033412}%
  \BibitemOpen
  \bibfield  {author} {\bibinfo {author} {\bibfnamefont {J.~E.}\ \bibnamefont
  {Johnson}}\ and\ \bibinfo {author} {\bibfnamefont {S.~L.}\ \bibnamefont
  {Rolston}},\ }\bibfield  {title} {\bibinfo {title} {Interactions between
  {R}ydberg-dressed atoms},\ }\href
  {https://doi.org/10.1103/PhysRevA.82.033412} {\bibfield  {journal} {\bibinfo
  {journal} {Phys. Rev. A}\ }\textbf {\bibinfo {volume} {82}},\ \bibinfo
  {pages} {033412} (\bibinfo {year} {2010})}\BibitemShut {NoStop}%
\bibitem [{\citenamefont {Khazali}\ \emph {et~al.}(2016)\citenamefont
  {Khazali}, \citenamefont {Lau}, \citenamefont {Humeniuk},\ and\ \citenamefont
  {Simon}}]{PhysRevA.94.023408}%
  \BibitemOpen
  \bibfield  {author} {\bibinfo {author} {\bibfnamefont {M.}~\bibnamefont
  {Khazali}}, \bibinfo {author} {\bibfnamefont {H.~W.}\ \bibnamefont {Lau}},
  \bibinfo {author} {\bibfnamefont {A.}~\bibnamefont {Humeniuk}},\ and\
  \bibinfo {author} {\bibfnamefont {C.}~\bibnamefont {Simon}},\ }\bibfield
  {title} {\bibinfo {title} {Large energy superpositions via {R}ydberg
  dressing},\ }\href {https://doi.org/10.1103/PhysRevA.94.023408} {\bibfield
  {journal} {\bibinfo  {journal} {Phys. Rev. A}\ }\textbf {\bibinfo {volume}
  {94}},\ \bibinfo {pages} {023408} (\bibinfo {year} {2016})}\BibitemShut
  {NoStop}%
\bibitem [{\citenamefont {Zeiher}\ \emph {et~al.}(2016)\citenamefont {Zeiher},
  \citenamefont {Van~Bijnen}, \citenamefont {Schau{\ss}}, \citenamefont {Hild},
  \citenamefont {Choi}, \citenamefont {Pohl}, \citenamefont {Bloch},\ and\
  \citenamefont {Gross}}]{zeiher2016}%
  \BibitemOpen
  \bibfield  {author} {\bibinfo {author} {\bibfnamefont {J.}~\bibnamefont
  {Zeiher}}, \bibinfo {author} {\bibfnamefont {R.}~\bibnamefont {Van~Bijnen}},
  \bibinfo {author} {\bibfnamefont {P.}~\bibnamefont {Schau{\ss}}}, \bibinfo
  {author} {\bibfnamefont {S.}~\bibnamefont {Hild}}, \bibinfo {author}
  {\bibfnamefont {J.-y.}\ \bibnamefont {Choi}}, \bibinfo {author}
  {\bibfnamefont {T.}~\bibnamefont {Pohl}}, \bibinfo {author} {\bibfnamefont
  {I.}~\bibnamefont {Bloch}},\ and\ \bibinfo {author} {\bibfnamefont
  {C.}~\bibnamefont {Gross}},\ }\bibfield  {title} {\bibinfo {title} {Many-body
  interferometry of a {R}ydberg-dressed spin lattice},\ }\href@noop {}
  {\bibfield  {journal} {\bibinfo  {journal} {Nature Physics}\ }\textbf
  {\bibinfo {volume} {12}},\ \bibinfo {pages} {1095} (\bibinfo {year}
  {2016})}\BibitemShut {NoStop}%
\bibitem [{\citenamefont {Henkel}\ \emph {et~al.}(2010)\citenamefont {Henkel},
  \citenamefont {Nath},\ and\ \citenamefont {Pohl}}]{PhysRevLett.104.195302}%
  \BibitemOpen
  \bibfield  {author} {\bibinfo {author} {\bibfnamefont {N.}~\bibnamefont
  {Henkel}}, \bibinfo {author} {\bibfnamefont {R.}~\bibnamefont {Nath}},\ and\
  \bibinfo {author} {\bibfnamefont {T.}~\bibnamefont {Pohl}},\ }\bibfield
  {title} {\bibinfo {title} {Three-dimensional roton excitations and supersolid
  formation in {R}ydberg-excited {B}ose-{E}instein condensates},\ }\href
  {https://doi.org/10.1103/PhysRevLett.104.195302} {\bibfield  {journal}
  {\bibinfo  {journal} {Phys. Rev. Lett.}\ }\textbf {\bibinfo {volume} {104}},\
  \bibinfo {pages} {195302} (\bibinfo {year} {2010})}\BibitemShut {NoStop}%
\bibitem [{\citenamefont {Bouchoule}\ and\ \citenamefont
  {Mølmer}(2002)}]{PhysRevA.65.041803}%
  \BibitemOpen
  \bibfield  {author} {\bibinfo {author} {\bibfnamefont {I.}~\bibnamefont
  {Bouchoule}}\ and\ \bibinfo {author} {\bibfnamefont {K.}~\bibnamefont
  {Mølmer}},\ }\bibfield  {title} {\bibinfo {title} {Spin squeezing of atoms
  by the dipole interaction in virtually excited {R}ydberg states},\ }\href
  {https://doi.org/10.1103/PhysRevA.65.041803} {\bibfield  {journal} {\bibinfo
  {journal} {Phys. Rev. A}\ }\textbf {\bibinfo {volume} {65}} (\bibinfo {year}
  {2002})}\BibitemShut {NoStop}%
\bibitem [{\citenamefont {Saffman}(2016)}]{Saffman_2016}%
  \BibitemOpen
  \bibfield  {author} {\bibinfo {author} {\bibfnamefont {M.}~\bibnamefont
  {Saffman}},\ }\bibfield  {title} {\bibinfo {title} {Quantum computing with
  atomic qubits and {R}ydberg interactions: progress and challenges},\ }\href
  {https://doi.org/10.1088/0953-4075/49/20/202001} {\bibfield  {journal}
  {\bibinfo  {journal} {Journal of Physics B: Atomic, Molecular and Optical
  Physics}\ }\textbf {\bibinfo {volume} {49}},\ \bibinfo {pages} {202001}
  (\bibinfo {year} {2016})}\BibitemShut {NoStop}%
\bibitem [{\citenamefont {Walker}\ and\ \citenamefont
  {Saffman}(2008)}]{PhysRevA.77.032723}%
  \BibitemOpen
  \bibfield  {author} {\bibinfo {author} {\bibfnamefont {T.~G.}\ \bibnamefont
  {Walker}}\ and\ \bibinfo {author} {\bibfnamefont {M.}~\bibnamefont
  {Saffman}},\ }\bibfield  {title} {\bibinfo {title} {Consequences of {Z}eeman
  degeneracy for the van der {W}aals blockade between {R}ydberg atoms},\ }\href
  {https://doi.org/10.1103/PhysRevA.77.032723} {\bibfield  {journal} {\bibinfo
  {journal} {Phys. Rev. A}\ }\textbf {\bibinfo {volume} {77}},\ \bibinfo
  {pages} {032723} (\bibinfo {year} {2008})}\BibitemShut {NoStop}%
\bibitem [{\citenamefont {Vaillant}\ \emph {et~al.}(2012)\citenamefont
  {Vaillant}, \citenamefont {Jones},\ and\ \citenamefont
  {Potvliege}}]{Vaillant_2012}%
  \BibitemOpen
  \bibfield  {author} {\bibinfo {author} {\bibfnamefont {C.~L.}\ \bibnamefont
  {Vaillant}}, \bibinfo {author} {\bibfnamefont {M.~P.~A.}\ \bibnamefont
  {Jones}},\ and\ \bibinfo {author} {\bibfnamefont {R.~M.}\ \bibnamefont
  {Potvliege}},\ }\bibfield  {title} {\bibinfo {title} {Long-range
  {R}ydberg{\textendash}{R}ydberg interactions in calcium, strontium and
  ytterbium},\ }\href {https://doi.org/10.1088/0953-4075/45/13/135004}
  {\bibfield  {journal} {\bibinfo  {journal} {Journal of Physics B: Atomic,
  Molecular and Optical Physics}\ }\textbf {\bibinfo {volume} {45}},\ \bibinfo
  {pages} {135004} (\bibinfo {year} {2012})}\BibitemShut {NoStop}%
\bibitem [{\citenamefont {Ding}\ \emph {et~al.}(2018)\citenamefont {Ding},
  \citenamefont {Whalen}, \citenamefont {Kanungo}, \citenamefont {Killian},
  \citenamefont {Dunning}, \citenamefont {Yoshida},\ and\ \citenamefont
  {Burgd\"orfer}}]{Ding2018}%
  \BibitemOpen
  \bibfield  {author} {\bibinfo {author} {\bibfnamefont {R.}~\bibnamefont
  {Ding}}, \bibinfo {author} {\bibfnamefont {J.~D.}\ \bibnamefont {Whalen}},
  \bibinfo {author} {\bibfnamefont {S.~K.}\ \bibnamefont {Kanungo}}, \bibinfo
  {author} {\bibfnamefont {T.~C.}\ \bibnamefont {Killian}}, \bibinfo {author}
  {\bibfnamefont {F.~B.}\ \bibnamefont {Dunning}}, \bibinfo {author}
  {\bibfnamefont {S.}~\bibnamefont {Yoshida}},\ and\ \bibinfo {author}
  {\bibfnamefont {J.}~\bibnamefont {Burgd\"orfer}},\ }\bibfield  {title}
  {\bibinfo {title} {Spectroscopy of $^{87}\mathrm{Sr}$ triplet {R}ydberg
  states},\ }\href@noop {} {\bibfield  {journal} {\bibinfo  {journal} {Phys.
  Rev. A}\ }\textbf {\bibinfo {volume} {98}},\ \bibinfo {pages} {042505}
  (\bibinfo {year} {2018})}\BibitemShut {NoStop}%
\bibitem [{\citenamefont {Robicheaux}\ \emph {et~al.}(2018)\citenamefont
  {Robicheaux}, \citenamefont {Booth},\ and\ \citenamefont
  {Saffman}}]{Robicheaux2018}%
  \BibitemOpen
  \bibfield  {author} {\bibinfo {author} {\bibfnamefont {F.}~\bibnamefont
  {Robicheaux}}, \bibinfo {author} {\bibfnamefont {D.~W.}\ \bibnamefont
  {Booth}},\ and\ \bibinfo {author} {\bibfnamefont {M.}~\bibnamefont
  {Saffman}},\ }\bibfield  {title} {\bibinfo {title} {Theory of long range
  interactions for {R}ydberg states attached to hyperfine split cores},\
  }\href@noop {} {\bibfield  {journal} {\bibinfo  {journal} {Phys. Rev. A}\
  }\textbf {\bibinfo {volume} {97}},\ \bibinfo {pages} {022508} (\bibinfo
  {year} {2018})}\BibitemShut {NoStop}%
\bibitem [{\citenamefont {Robicheaux}(2019)}]{Robicheaux2019}%
  \BibitemOpen
  \bibfield  {author} {\bibinfo {author} {\bibfnamefont {F.}~\bibnamefont
  {Robicheaux}},\ }\bibfield  {title} {\bibinfo {title} {Calculations of long
  range interactions for ${}^{87}${S}r {R}ydberg states},\ }\href@noop {}
  {\bibfield  {journal} {\bibinfo  {journal} {J. Phys. B: At. Mol. Opt. Phys.}\
  }\textbf {\bibinfo {volume} {52}},\ \bibinfo {pages} {244001} (\bibinfo
  {year} {2019})}\BibitemShut {NoStop}%
\bibitem [{\citenamefont {Jau}\ \emph {et~al.}(2016)\citenamefont {Jau},
  \citenamefont {Hankin}, \citenamefont {Keating}, \citenamefont {Deutsch},\
  and\ \citenamefont {Biedermann}}]{jau2016}%
  \BibitemOpen
  \bibfield  {author} {\bibinfo {author} {\bibfnamefont {Y.-Y.}\ \bibnamefont
  {Jau}}, \bibinfo {author} {\bibfnamefont {A.}~\bibnamefont {Hankin}},
  \bibinfo {author} {\bibfnamefont {T.}~\bibnamefont {Keating}}, \bibinfo
  {author} {\bibfnamefont {I.}~\bibnamefont {Deutsch}},\ and\ \bibinfo {author}
  {\bibfnamefont {G.}~\bibnamefont {Biedermann}},\ }\bibfield  {title}
  {\bibinfo {title} {Entangling atomic spins with a {R}ydberg-dressed spin-flip
  blockade},\ }\href@noop {} {\bibfield  {journal} {\bibinfo  {journal} {Nature
  Physics}\ }\textbf {\bibinfo {volume} {12}},\ \bibinfo {pages} {71} (\bibinfo
  {year} {2016})}\BibitemShut {NoStop}%
\bibitem [{\citenamefont {Goldschmidt}\ \emph {et~al.}(2016)\citenamefont
  {Goldschmidt}, \citenamefont {Boulier}, \citenamefont {Brown}, \citenamefont
  {Koller}, \citenamefont {Young}, \citenamefont {Gorshkov}, \citenamefont
  {Rolston},\ and\ \citenamefont {Porto}}]{goldschmidt2016}%
  \BibitemOpen
  \bibfield  {author} {\bibinfo {author} {\bibfnamefont {E.~A.}\ \bibnamefont
  {Goldschmidt}}, \bibinfo {author} {\bibfnamefont {T.}~\bibnamefont
  {Boulier}}, \bibinfo {author} {\bibfnamefont {R.~C.}\ \bibnamefont {Brown}},
  \bibinfo {author} {\bibfnamefont {S.~B.}\ \bibnamefont {Koller}}, \bibinfo
  {author} {\bibfnamefont {J.~T.}\ \bibnamefont {Young}}, \bibinfo {author}
  {\bibfnamefont {A.~V.}\ \bibnamefont {Gorshkov}}, \bibinfo {author}
  {\bibfnamefont {S.}~\bibnamefont {Rolston}},\ and\ \bibinfo {author}
  {\bibfnamefont {J.~V.}\ \bibnamefont {Porto}},\ }\bibfield  {title} {\bibinfo
  {title} {Anomalous broadening in driven dissipative {R}ydberg systems},\
  }\href@noop {} {\bibfield  {journal} {\bibinfo  {journal} {Phys. Rev. Let.}\
  }\textbf {\bibinfo {volume} {116}},\ \bibinfo {pages} {113001} (\bibinfo
  {year} {2016})}\BibitemShut {NoStop}%
\bibitem [{\citenamefont {Boulier}\ \emph {et~al.}(2017)\citenamefont
  {Boulier}, \citenamefont {Magnan}, \citenamefont {Bracamontes}, \citenamefont
  {Maslek}, \citenamefont {Goldschmidt}, \citenamefont {Young}, \citenamefont
  {Gorshkov}, \citenamefont {Rolston},\ and\ \citenamefont
  {Porto}}]{boulier2017}%
  \BibitemOpen
  \bibfield  {author} {\bibinfo {author} {\bibfnamefont {T.}~\bibnamefont
  {Boulier}}, \bibinfo {author} {\bibfnamefont {E.}~\bibnamefont {Magnan}},
  \bibinfo {author} {\bibfnamefont {C.}~\bibnamefont {Bracamontes}}, \bibinfo
  {author} {\bibfnamefont {J.}~\bibnamefont {Maslek}}, \bibinfo {author}
  {\bibfnamefont {E.}~\bibnamefont {Goldschmidt}}, \bibinfo {author}
  {\bibfnamefont {J.}~\bibnamefont {Young}}, \bibinfo {author} {\bibfnamefont
  {A.~V.}\ \bibnamefont {Gorshkov}}, \bibinfo {author} {\bibfnamefont
  {S.}~\bibnamefont {Rolston}},\ and\ \bibinfo {author} {\bibfnamefont {J.~V.}\
  \bibnamefont {Porto}},\ }\bibfield  {title} {\bibinfo {title} {Spontaneous
  avalanche dephasing in large {R}ydberg ensembles},\ }\href@noop {} {\bibfield
   {journal} {\bibinfo  {journal} {Phys. Rev. A}\ }\textbf {\bibinfo {volume}
  {96}},\ \bibinfo {pages} {053409} (\bibinfo {year} {2017})}\BibitemShut
  {NoStop}%
\bibitem [{\citenamefont {Barredo}\ \emph {et~al.}(2018)\citenamefont
  {Barredo}, \citenamefont {Lienhard}, \citenamefont {de~L{\'e}s{\'e}leuc},
  \citenamefont {Lahaye},\ and\ \citenamefont {Browaeys}}]{Barredo2018}%
  \BibitemOpen
  \bibfield  {author} {\bibinfo {author} {\bibfnamefont {D.}~\bibnamefont
  {Barredo}}, \bibinfo {author} {\bibfnamefont {V.}~\bibnamefont {Lienhard}},
  \bibinfo {author} {\bibfnamefont {S.}~\bibnamefont {de~L{\'e}s{\'e}leuc}},
  \bibinfo {author} {\bibfnamefont {T.}~\bibnamefont {Lahaye}},\ and\ \bibinfo
  {author} {\bibfnamefont {A.}~\bibnamefont {Browaeys}},\ }\bibfield  {title}
  {\bibinfo {title} {Synthetic three-dimensional atomic structures assembled
  atom by atom},\ }\href {https://doi.org/10.1038/s41586-018-0450-2} {\bibfield
   {journal} {\bibinfo  {journal} {Nature}\ }\textbf {\bibinfo {volume}
  {561}},\ \bibinfo {pages} {79} (\bibinfo {year} {2018})}\BibitemShut
  {NoStop}%
\end{thebibliography}%


\providecommand{\noopsort}[1]{}\providecommand{\singleletter}[1]{#1}%
%

\end{document}